\documentclass[10pt]{article}
\usepackage{graphicx}
\usepackage{amsmath}
\usepackage{amssymb}
\usepackage{caption2}
\begin{document}
\bibliographystyle{prsty}
\begin{center}
{\large {\bf \sc{  Analysis of  the hidden-charm pentaquark candidates in the $J/\psi \Xi^*$ mass spectrum  via the  QCD sum rules }}} \\[2mm]
Zhi-Gang Wang \footnote{E-mail: zgwang@aliyun.com.  }  \\
 Department of Physics, North China Electric Power University, Baoding 071003, P. R. China
\end{center}

\begin{abstract}
In this work, we explore  the diquark-diquark-antiquark type decuplet
  hidden-charm pentaquark  states with the symbolic valence  structure $qssc\bar{c}$  via  the QCD sum rules  extensively, and
 achieve  the spectroscopy of the lowest decuplet $qssc\bar{c}$
   pentaquark states with the quantum numbers  $IJ^{P}=\frac{1}{2}{\frac{1}{2}}^-$, $\frac{1}{2}{\frac{3}{2}}^-$ and $\frac{1}{2}{\frac{5}{2}}^-$, and suggest to explore these exotic states   in the exclusive  processes
$\Xi_b^{\prime0}
\to P_{css}^0\,\phi \to J/\psi \Xi^{*0}  \phi $ and
$\Omega_b^{-}\to P_{css}^-\, \bar{K}^0 \to J/\psi \Xi^{*-}\, \bar{K}^0$. As a byproduct, we can re-testify   classifications of the light baryons by investigating  the hidden-charm pentaquark  decays.
\end{abstract}



\section{Introduction}
In 1964, Gell-Mann suggested  that  multiquark states beyond the  minimal valence quark  constituents  $q\bar{q}$ and $qqq$  might exist \cite{Gell-Mann-1964}. Latter, the five-quark baryons with the quark constituents $qqqq\bar{q}$ were developed \cite{Strottman-1979}, and the name pentaquark  was introduced by Lipkin \cite{Lipkin-1987}.
A diquark  can effectively serve  as an antiquark, thus two diquarks and one antiquark can attract each
other by the same means as three antiquarks do in an ordinary antibaryon.

Although some
pentaquark candidates  were claimed in the experiments before 2015, none of them
survived scrutiny of additional experimental data, for example,
a candidate for a $K^+n$ resonance called $\Theta^+(1540)$, whose minimal quark
constituents  would be $uudd\bar{s}$, was claimed in 2003
\cite{exp2003-Cita1540-1,exp2003-Cita1540-2,exp2003-Cita1540-3},
a candidate for a $D^{*\pm} p^{\mp}$ resonance  called $\Theta_c^0(3099)$, whose minimal quark
constituents would be $uudd\bar{c}$, was claimed in 2004  \cite{exp2004-cuudd}.

In 2015,  the  LHCb collaboration  observed  two exotic structures $P_c(4380)$ and $P_c(4450)$ with  preferred spin-parity   $J^P={\frac{3}{2}}^-$ and ${\frac{5}{2}}^+$ respectively in the $J/\psi p$  mass spectrum in the exclusive process $\Lambda_b^0\to J/\psi pK^- $ \cite{LHCb-4380}, and stimulated  deep interest in the spectroscopy of hidden-charm pentaquark states and opened a new era for  deeper  understanding the basic subjects, such as confinement, dynamically symmetry breaking, etc.
In 2019, the LHCb collaboration re-investigated  the same process but a data sample with an order of magnitude larger than previously one, confirmed the old structure $P_c(4450)$, which  is made up of two narrow superimposed  structures $P_c(4440)$ and $P_c(4457)$, and observed a new structure  $P_c(4312)$ \cite{LHCb-Pc4312}. If these  $P_c$ states are genuine  resonances, not kinematical effects, such as threshold cusps, triangle singularities, coupled channel effects, etc, the minimal valence quark constituents should be $uudc\bar{c}$, irrespective of the diquark-diquark-antiquark type, diquark-triquark type  or meson-baryon type \cite{Review-penta-mole-ZhSL-RPT,Review-penta-Esposito-RPT,
Review-penta-Ali-PPNP,Review-penta-mole-GuoFK-RMP,
Review-penta-mole-LiuYR-PPNP,Review-penta-mole-Brambilla-RPT}.

In 2020, the LHCb collaboration investigated the exclusive process $\Xi_b^- \to J/\psi K^- \Lambda$, and discovered   a hidden-charm structure  $P_{cs}(4459)$ with strangeness for the first time in the $J/\psi \Lambda$ mass spectrum with  a  significance of  $3.1\sigma$ \cite{LHCb-Pcs4459-2012}.

In 2021, the LHCb collaboration investigated the exclusive process $B_s^0 \to J/\psi p\bar{p}$, and  discovered  a hidden-charm structure $P_c(4337)$ without strangeness in the $J/\psi p$ and $J/\psi \bar{p}$ systems  with a significance approximately  3.1 to 3.7$\sigma$  according to the possible assumed spin-parity \cite{LHCb-Pc4337}.
In 2022, the LHCb collaboration investigated the exclusive process $B^- \to J/\psi \Lambda \bar{p}$ and discovered  a new hidden-charm structure $P_{cs}(4338)$ with strangeness  in the $J/\psi \Lambda$ mass spectrum with the favored  spin-parity $J^P={\frac{1}{2}}^-$ \cite{LHCb-Pcs4338}.
Compared to the $\Lambda_b$ decays, the $B$-decays are not optimal channels to investigate the exotic $P$ states due to additional quark-antiquark pairs have to be created from the QCD vacuum, and the corresponding branching fractions are expected to be smaller.

In 2025, the Belle and Belle-II collaborations observed the $\Upsilon(1{\rm S }, 2{\rm S})$ inclusive decays to the final states $J/\psi\Lambda(\bar{\Lambda})$ for the first time, and discovered   an evidence of (just an evidence, not confirmed) the old structure $P_{cs}(4459)$  with a local significance of $3.3\,\sigma$ \cite{Belle-Pcs4338-Pcs4459}. Again, if these $P_{cs}$ states are genuine  resonances, the minimal valence quark constituents should be $udsc\bar{c}$, irrespective of the diquark-diquark-antiquark type, diquark-triquark-type or meson-baryon type \cite{Review-mole-WangB-PRT,Review-mole-GengLS-PRT,WangZG-Review}.

On the other hand, the  LHCb collaboration observed the exclusive  decays
 $\Lambda_{b}^{0} \to \Lambda_{c}^{+} \bar{D}^{0} K^{-}$, $\Lambda_{c}^{+} \bar{D}^{*0} K^{-}$  \cite{LHCb-LamcD-EPJC-2024},
 $ \Sigma_c^{*++} D^{*-} K^{-}$, $ \Sigma_c^{*++} D^{-} K^{-}$, $ \Sigma_c^{++} D^{*-} K^{-}$, $ \Sigma_c^{++} D^{-} K^{-}$
\cite{LHCb-SigmaD-PRD-2024}, $\Lambda_{c}^{+}D_{s}^{-}K^{+}K^{-}$ \cite{LHCb-LamcDs-PRD-2025}, $ J/\psi\Xi^- K^+$ and $\Xi_b^0 \to J/\psi\Xi^- \pi^+$ \cite{LHCb-JpsiXi-EPJC-2025}, but discovered no evidence for both the $P_c$ and $P_{cs}$ states. However, we cannot exclude the possibility that when more experimental data are accumulated, the exotic states appear.

These decays occur through the CKM favored process $b \to c\bar{c}s$ (via emission of a virtual $W$ meson) at the quark level with the weak effective Hamiltonian $H_w$,
\begin{eqnarray}
H_w &=& \frac{G_F}{\sqrt{2}}\,V_{c b} V_{c s}^*
\Big[ \widetilde{C}_1\, \overline{c}\Gamma_\mu c\,
\overline{s}\Gamma^\mu b +  \widetilde{C}_2\, \overline{c}\Gamma_\mu\frac{\lambda^a}{2} c\,
\overline{s}\Gamma^\mu\frac{\lambda^a}{2} b\Big]   \, ,
\end{eqnarray}
where $\Gamma_\mu=\gamma_\mu(1-\gamma_5)$, the $G_F$ is the Fermi coupling constant, the $V_{cb/cs}$ are the CKM matrix elements,  the $\widetilde{C}_{1/2}$ are the Wilson coefficients \cite{AJBuras-1996}. For the $\Lambda$-type baryons $\Lambda_b$ and $\Xi_b$,
 the diquarks $ud$ and $qs$ are in the light-flavor antitriplet $\bar{\mathbf{3}}$ representation, while the $H_w$ is in the light-flavor triplet $\mathbf{3}$ representation, the decay-chains,
\begin{eqnarray}
\Lambda_b^0(bud)&\to& P_c^+ (\mathbf{8})\,K^- \to J/\psi p \,K^-\, ,\nonumber\\
&\to& P_{cs}^0(\mathbf{8})\,\phi \to J/\psi \Lambda^0\, \phi \, ,
\end{eqnarray}
\begin{eqnarray}
\Xi_b^-(bds)&\to& P_{cs}^0\,(\mathbf{8}) K^-\to J/\psi \Lambda^0\, K^-\, ,\nonumber\\
&\to&P_{css}^-(\mathbf{8})\,\phi\to J/\psi \Xi^-  \phi \, ,
\end{eqnarray}
are optimal channels to investigate the hidden-charm pentaquark states in the octet $\mathbf{8}$  representation.
For the  $\Sigma$-type baryons $\Sigma_b$, $\Xi^\prime_b$ and $\Omega_b$,
 the diquarks $qq$, $qs$ and $ss$ are in the light-flavor sextet $\mathbf{6}$ representation, the decay-chains,
\begin{eqnarray}
\Sigma_b^+(buu)&\to& P_c^{++} (\mathbf{10})\,K^- \to J/\psi \Delta^{++} \,K^-\, ,\nonumber\\
&\to& P_{cs}^+(\mathbf{10})\,\phi \to J/\psi \Sigma^{*+}\, \phi \, ,\nonumber\\
&\to& P_{cs}^+(\mathbf{8})\,\phi \to J/\psi \Sigma^+\, \phi \, ,
\end{eqnarray}
\begin{eqnarray}
\Xi_b^{\prime0}(bus)&\to& P_{cs}^+\,(\mathbf{10}) K^-\to J/\psi \Sigma^{*+}\, K^-\, ,\nonumber\\
&\to& P_{cs}^+\,(\mathbf{8}) K^- \to J/\psi \Sigma^{+}\, K^-\, ,\nonumber\\
&\to&P_{css}^0(\mathbf{10})\,\phi \to J/\psi \Xi^{*0}  \phi \, , \nonumber\\
&\to&P_{css}^0(\mathbf{8})\,\phi \to J/\psi \Xi^0  \phi \, ,
\end{eqnarray}
\begin{eqnarray}
\Omega_b^{-}(bss)&\to& P_{css}^-\,(\mathbf{10}) \bar{K}^0 \to J/\psi \Xi^{*-}\, \bar{K}^0\, ,\nonumber\\
&\to& P_{css}^-\,(\mathbf{8}) \bar{K}^0 \to J/\psi \Xi^{-}\, \bar{K}^0\, ,\nonumber\\
&\to&P_{csss}^-(\mathbf{10})\,\phi \to J/\psi \Omega^{-}  \phi \, ,
\end{eqnarray}
are optimal channels to investigate the hidden-charm pentaquark states in the octet $\mathbf{8}$ and decuplet $\mathbf{10}$ representations \cite{di-di-anti-penta-1,di-di-anti-penta-2}. In the scenario of molecular states, the spectroscopy and production mechanism are quite different \cite{Oset-1,Oset-2}.

These pentaquark candidates usually  lie close to the meson-baryon  thresholds except for the $P_c(4337)$,
 which results in the natural interpretations  of loose molecular states. In the case of the exotic mesons, only few of them lie just close to the meson-meson thresholds, we have to turn to large coupled-channel effects  to outcome the discrepancies.  If we restrict to the method of the QCD sum rules,
 we could reproduce the masses of the existing $P_c$ and $P_{cs}$ states in the scenario of  color singlet-singlet type pentaquark states  except for the $P_c(4337)$ \cite{WangZG-IJMPA-3-scheme,Pcs4459-mole-WangZG-SR,Pc4312-mole-penta-WXW-SCPMA,
Pc4312-mole-penta-WXW-IJMPA,Pcs4338-mole-XWWang,QCDSR-penta-mole-ChenHX-PRD-2019,QCDSR-penta-mole-ChenHX-EPJC-2021,
QCDSR-penta-mole-Azizi-PRD-2017,QCDSR-penta-mole-ZhangJR-EPJC-2019,
QCDSR-penta-mole-Azizi-PRD-2021}.  While in the scenario of  diquark-diquark-antiquark type  pentaquark states, there are enough rooms to accommodate  all the observed $P$ states, such as the $P_c(4312)$, $P_c(4337)$, $P_{cs}(4338)$, $P_{c}(4380)$, $P_c(4440)$, $P_c(4457)$, $P_{cs}(4459)$ \cite{WangZG-Pc12-JpsiLambda,WangZG-Pc12-Jpsip}.

For the light-heavy systems having emblematic  quark constituents  $qqqQ\bar{Q}$, where $q=u$, $d$, $s$, $Q=b$, $c$, we could distinguish light and heavy degrees of freedoms plainly, the components $qqq$ have the light-flavor $SU(3)$ symmetry and the components $Q\bar{Q}$ have the heavy quark symmetry. Thus,
\begin{eqnarray}\label{two-octet}
{\mathbf{3}}\otimes {\mathbf{3}}\otimes {\mathbf{3}} &\to & \left({\bar{\mathbf{3}}} \oplus {\mathbf{6}} \right) \otimes {\mathbf{3}}\, , \nonumber\\
&\to& {\mathbf{1}} \oplus {\mathbf{8}}_1 \oplus {\mathbf{8}}_2\oplus {\mathbf{10}}\, ,
\end{eqnarray}
there are one singlet, two octets and one decuplet hidden-charm pentaquark states.
In Refs.\cite{Wang1508-EPJC,WangHuang-EPJC-1508-12,WangZG-EPJC-1509-12,
 WangZG-NPB-1512-32,WangZhang-APPB}, we investigated the diquark-diquark-antiquark type hidden-charm pentaquark states with the spin-parity  $J^P={\frac{1}{2}}^\pm$, ${\frac{3}{2}}^\pm$, ${\frac{5}{2}}^\pm$  and without strangeness, with single strangeness, with double strangeness and with triple strangeness   via   the QCD sum rules  extensively by fulfilling  the operator product expansion (OPE) up to the vacuum condensates of dimension 10, but did not exhaust all the lowest quark configurations.

In Refs.\cite{WangZG-Pc12-JpsiLambda,WangZG-Pc12-Jpsip,WZG-penta-IJMPA,WangZG-Pc12-JpsiXi,WangZG-Pc12-JpsiSgm}, we updated our  old calculations by fulfilling the OPE up to the vacuum condensates of dimension 13, tried to  exhaust all the lowest hidden-charm five-quark configurations, and comprehensively investigated  the octet and decuplet hidden-charm
pentaquark states expected to have the typical decay channels  $J/\psi \Lambda$, $J/\psi p$, $J/\psi \Xi$, $J/\psi \Sigma$, $J/\psi \Delta$ respectively   with the QCD sum rules, and made reliable predications for the spectroscopy.

Now we update our old calculations on the decuplet hidden-charm pentaquark state expected to have the typical decay channels $J/\psi \Xi^*$   \cite{WangZG-EPJC-1509-12,WangZG-NPB-1512-32},  and try to exhaust all the lowest five-quark configurations.

 The article is arranged as follows:  we obtain the QCD sum rules for  the decuplet $qssc\bar{c}$ states in Sect.2;  in Sect.3, we present the numerical results and discussions; and Sect.4 is reserved for our
conclusion.

\section{QCD sum rules for  the  $qssc\bar{c}$ pentaquark states}
As usual, we adopt the  correlation functions $\Pi(p)$, $\Pi_{\mu\nu}(p)$ and $\Pi_{\mu\nu\alpha\beta}(p)$,
\begin{eqnarray}\label{CF-Pi-Pi-Pi}
\Pi(p)&=&i\int d^4x e^{ip \cdot x} \langle0|T\left\{J(x)\bar{J}(0)\right\}|0\rangle \, ,\nonumber\\
\Pi_{\mu\nu}(p)&=&i\int d^4x e^{ip \cdot x} \langle0|T\left\{J_{\mu}(x)\bar{J}_{\nu}(0)\right\}|0\rangle \, ,\nonumber\\
\Pi_{\mu\nu\alpha\beta}(p)&=&i\int d^4x e^{ip \cdot x} \langle0|T\left\{J_{\mu\nu}(x)\bar{J}_{\alpha\beta}(0)\right\}|0\rangle \, ,
\end{eqnarray}
where
 \begin{eqnarray}
 J(x)&=&J^1(x)\, , \, J^2(x)\, , \nonumber\\
 J_\mu(x)&=&J_\mu^1(x)\, , \, J_\mu^2(x)\, , \, J_\mu^3(x)\, ,  \nonumber\\
 J_{\mu\nu}(x)&=&J_{\mu\nu}^1(x)\, , \, J_{\mu\nu}^2(x)\, ,
 \end{eqnarray}
 and
\begin{eqnarray}\label{Current-12}
J^{1}(x)&=&\frac{\varepsilon^{ila} \varepsilon^{ijk}\varepsilon^{lmn}}{\sqrt{3}} \left[ s^T_j(x) C\gamma_\mu s_k(x)q^T_m(x) C\gamma^\mu c_n(x)+2s^T_j(x) C\gamma_\mu q_k(x)s^T_m(x) C\gamma^\mu c_n(x) \right]  C\bar{c}^{T}_{a}(x) \, , \nonumber\\
J^{2}(x)&=&\frac{\varepsilon^{ila} \varepsilon^{ijk}\varepsilon^{lmn}}{\sqrt{3}} \left[ s^T_j(x) C\gamma_\mu s_k(x) q^T_m(x) C\gamma_5 c_n(x)+2s^T_j(x) C\gamma_\mu q_k(x) s^T_m(x) C\gamma_5 c_n(x)\right] \gamma_5 \gamma^\mu  C\bar{c}^{T}_{a}(x)
 \, , \nonumber \\
 \end{eqnarray}
 with the isospin-spin $(I,J)=(\frac{1}{2},\frac{1}{2})$ \cite{WangZG-EPJC-1509-12},
\begin{eqnarray}\label{Current-32}
 J^{1}_{\mu}(x)&=&\frac{\varepsilon^{ila} \varepsilon^{ijk}\varepsilon^{lmn}}{\sqrt{3}} \left[ s^T_j(x) C\gamma_\mu s_k(x) q^T_m(x) C\gamma_5 c_n(x) +2s^T_j(x) C\gamma_\mu q_k(x) s^T_m(x) C\gamma_5 c_n(x)\right]   C\bar{c}^{T}_{a}(x) \, , \nonumber \\
 J^{2}_{\mu}(x)&=&\frac{\varepsilon^{ila} \varepsilon^{ijk}\varepsilon^{lmn}}{\sqrt{3}} \left[ s^T_j(x) C\gamma_\mu s_k(x)q^T_m(x) C\gamma_\alpha c_n(x)+2s^T_j(x) C\gamma_\mu q_k(x)s^T_m(x) C\gamma_\alpha c_n(x) \right] \gamma_5\gamma^\alpha C\bar{c}^{T}_{a}(x) \, , \nonumber\\
J^{3}_{\mu}(x)&=&\frac{\varepsilon^{ila} \varepsilon^{ijk}\varepsilon^{lmn}}{\sqrt{3}} \left[ s^T_j(x) C\gamma_\alpha s_k(x)q^T_m(x) C\gamma_\mu c_n(x)+2s^T_j(x) C\gamma_\alpha q_k(x)s^T_m(x) C\gamma_\mu c_n(x) \right] \gamma_5\gamma^\alpha C\bar{c}^{T}_{a}(x) \, ,\nonumber\\
 \end{eqnarray}
 with the isospin-spin $(I,J)=(\frac{1}{2},\frac{3}{2})$ \cite{WangZG-NPB-1512-32},
\begin{eqnarray} \label{Current-52}
J^1_{\mu\nu}(x)&=&\frac{\varepsilon^{ila} \varepsilon^{ijk}\varepsilon^{lmn} }{\sqrt{6}}\, s^T_j(x) C\gamma_\mu s_k(x)\, q^T_m(x) C\gamma_5 c_n(x)\, \gamma_5\gamma_{\nu}C\bar{c}^{T}_{a}(x)+(\mu \leftrightarrow \nu) \, ,\nonumber\\
&&+\frac{\varepsilon^{ila} \varepsilon^{ijk}\varepsilon^{lmn} }{\sqrt{6}}\, 2s^T_j(x) C\gamma_\mu q_k(x)\, s^T_m(x) C\gamma_5 c_n(x)\, \gamma_5\gamma_{\nu}C\bar{c}^{T}_{a}(x)+(\mu \leftrightarrow \nu) \, ,\nonumber\\
J^2_{\mu\nu}(x)&=&\frac{\varepsilon^{ila} \varepsilon^{ijk}\varepsilon^{lmn}}{\sqrt{6}} s^T_j(x) C\gamma_\mu s_k(x)\, q^T_m(x) C\gamma_\nu c_n(x)  C\bar{c}^{T}_{a}(x)+(\mu \leftrightarrow \nu)\, ,\nonumber\\
&&+\frac{\varepsilon^{ila} \varepsilon^{ijk}\varepsilon^{lmn}}{\sqrt{6}} 2s^T_j(x) C\gamma_\mu q_k(x)\, s^T_m(x) C\gamma_\nu c_n(x)  C\bar{c}^{T}_{a}(x)+(\mu \leftrightarrow \nu)\, ,
\end{eqnarray}
with the isospin-spin $(I,J)=(\frac{1}{2},\frac{5}{2})$,
$q=u$, $d$, the subscripts $i$, $j$, $k$, $l$, $m$, $n$ and $a$ are color indexes.

The constituents  $\varepsilon^{ijk}q^T_jC\gamma_{\mu}s_k$ and  $\varepsilon^{ijk}s^T_jC\gamma_{\mu}s_k$ have spins $S_L=1$, the constituents $\varepsilon^{ijk}q^T_jC\gamma_5c_k$, $\varepsilon^{ijk}s^T_jC\gamma_5c_k$, $\varepsilon^{ijk}q^T_jC\gamma_{\mu}c_k$ and   $\varepsilon^{ijk}s^T_jC\gamma_{\mu}c_k$ have spins $S_H=0$, $0$, $1$ and $1$, respectively, where the subscripts $L$ and $H$ symbolize  the light  and heavy diquarks respectively.  A heavy diquark attracts  a light  one to form a four-quark cluster  in the color triplet,  the  angular momentum $\vec{J}_{LH}=\vec{S}_L+\vec{S}_H$ has possible  values $J_{LH}=0$, $1$, $2$.
On the other hand, the operators $C\bar{c}_a^T$ and $\gamma_5\gamma_{\mu}C\bar{c}_a^T$ have the spin-parity $J^P={\frac{1}{2}}^-$ and ${\frac{3}{2}}^-$, respectively. Thus,  the total angular momentums   $\vec{J}=\vec{J}_{LH}+\vec{J}_{\bar{c}}$ have possible values $J=\frac{1}{2}$, $\frac{3}{2}$, $\frac{5}{2}$, which are illustrated  plainly  in Table \ref{current-pentaQ}. The three light quarks $qss$ are symmetric, thus they are in the decuplet representation.

\begin{table}
\begin{center}
\begin{tabular}{|c|c|c|c|c|c|c|c|c|}\hline\hline
$[qq][qc]\bar{c}$ ($S_L$, $S_H$, $J_{LH}$, $J$)  & $J^{P}$             & Currents              \\ \hline

$[ss][qc]\bar{c}+2[sq][sc]\bar{c}$ ($1$, $1$, $0$, $\frac{1}{2}$) &${\frac{1}{2}}^{-}$  &$J^1(x)$        \\

$[ss][qc]\bar{c}+2[sq][sc]\bar{c}$ ($1$, $0$, $1$, $\frac{1}{2}$) &${\frac{1}{2}}^{-}$  &$J^2(x)$         \\    \hline

$[ss][qc]\bar{c}+2[sq][sc]\bar{c}$ ($1$, $0$, $1$, $\frac{3}{2}$) &${\frac{3}{2}}^{-}$ &$J^1_\mu(x)$          \\

$[ss][qc]\bar{c}+2[sq][sc]\bar{c}$ ($1$, $1$, $2$, $\frac{3}{2}$)${}_2$ &${\frac{3}{2}}^{-}$  &$J^2_\mu(x)$   \\

$[ss][qc]\bar{c}+2[sq][sc]\bar{c}$ ($1$, $1$, $2$, $\frac{3}{2}$)${}_3$ &${\frac{3}{2}}^{-}$  &$J^3_\mu(x)$   \\ \hline

$[ss][qc]\bar{c}+2[sq][sc]\bar{c}$ ($1$, $0$, $1$, $\frac{5}{2}$) &${\frac{5}{2}}^{-}$  &$J^1_{\mu\nu}(x)$    \\

$[ss][qc]\bar{c}+2[sq][sc]\bar{c}$ ($1$, $1$, $2$, $\frac{5}{2}$) &${\frac{5}{2}}^{-}$  &$J^2_{\mu\nu}(x)$   \\
\hline\hline
\end{tabular}
\end{center}
\caption{ The valence quark structures and spin-parity of the  currents.  }\label{current-pentaQ}
\end{table}

The currents $J(x)$, $J_\mu(x)$ and $J_{\mu\nu}(x)$ have the spin-parity
$J^P={\frac{1}{2}}^-$, ${\frac{3}{2}}^-$ and ${\frac{5}{2}}^-$, respectively, and
 couple potentially to the decuplet $qssc\bar{c}$  pentaquark states (P) both with negative  and positive parities based on quark-hadron duality,  as multiplying $i \gamma_{5}$ to the currents   $J(x)$, $J_\mu(x)$ and $J_{\mu\nu}(x)$ changes their parities \cite{WangZG-Review,Wang1508-EPJC}. We write down the couplings plainly, \begin{eqnarray}\label{Coupling12}
\langle 0| J (0)|P_{\frac{1}{2}}^{-}(p)\rangle &=&\lambda^{-}_{\frac{1}{2}} U^{-}(p,s) \, , \nonumber \\
\langle 0| J (0)|P_{\frac{1}{2}}^{+}(p)\rangle &=&\lambda^{+}_{\frac{1}{2}} i\gamma_5 U^{+}(p,s) \, ,
\end{eqnarray}
\begin{eqnarray}
\langle 0| J_{\mu} (0)|P_{\frac{3}{2}}^{-}(p)\rangle &=&\lambda^{-}_{\frac{3}{2}} U^{-}_\mu(p,s) \, ,  \nonumber \\
\langle 0| J_{\mu} (0)|P_{\frac{3}{2}}^{+}(p)\rangle &=&\lambda^{+}_{\frac{3}{2}}i\gamma_5 U^{+}_\mu(p,s) \, ,  \nonumber \\
\langle 0| J_{\mu} (0)|P_{\frac{1}{2}}^{+}(p)\rangle &=&f^{+}_{\frac{1}{2}}p_\mu U^{+}(p,s) \, , \nonumber \\
\langle 0| J_{\mu} (0)|P_{\frac{1}{2}}^{-}(p)\rangle &=&f^{-}_{\frac{1}{2}}p_\mu i\gamma_5 U^{-}(p,s) \, ,
\end{eqnarray}
\begin{eqnarray}\label{Coupling52}
\langle 0| J_{\mu\nu} (0)|P_{\frac{5}{2}}^{-}(p)\rangle &=&\sqrt{2}\lambda^{-}_{\frac{5}{2}} U^{-}_{\mu\nu}(p,s) \, ,\nonumber\\
\langle 0| J_{\mu\nu} (0)|P_{\frac{5}{2}}^{+}(p)\rangle &=&\sqrt{2}\lambda^{+}_{\frac{5}{2}}i\gamma_5 U^{+}_{\mu\nu}(p,s) \, ,\nonumber\\
\langle 0| J_{\mu\nu} (0)|P_{\frac{3}{2}}^{+}(p)\rangle &=&f^{+}_{\frac{3}{2}} \left[p_\mu U^{+}_{\nu}(p,s)+p_\nu U^{+}_{\mu}(p,s)\right] \, , \nonumber\\
\langle 0| J_{\mu\nu} (0)|P_{\frac{3}{2}}^{-}(p)\rangle &=&f^{-}_{\frac{3}{2}}i\gamma_5 \left[p_\mu U^{-}_{\nu}(p,s)+p_\nu U^{-}_{\mu}(p,s)\right] \, , \nonumber\\
\langle 0| J_{\mu\nu} (0)|P_{\frac{1}{2}}^{-}(p)\rangle &=&g^{-}_{\frac{1}{2}}p_\mu p_\nu U^{-}(p,s) \, , \nonumber\\
\langle 0| J_{\mu\nu} (0)|P_{\frac{1}{2}}^{+}(p)\rangle &=&g^{+}_{\frac{1}{2}}p_\mu p_\nu i\gamma_5 U^{+}(p,s) \, ,
\end{eqnarray}
where  the $\mp$ are  the  parities, the  $\frac{1}{2}$, $\frac{3}{2}$ and $\frac{5}{2}$  are  the spins,   the $\lambda$, $f$ and $g$ are the pole residues, and the $U^\pm(p,s)$,  $U^{\pm}_\mu(p,s)$ and $U^{\pm}_{\mu\nu}(p,s)$ are Dirac and Rarita-Schwinger spinors respectively, one can consult
Refs.\cite{WangZG-Review,Wang1508-EPJC} for more details.

Based on above discussions, we insert  a complete set  of intermediate
decuplet $qssc\bar{c}$  pentaquark states with the same quantum numbers as the currents  $J(x)$, $i\gamma_5 J(x)$, $J_{\mu}(x)$, $i\gamma_5 J_{\mu}(x)$, $J_{\mu\nu}(x)$  and $i\gamma_5 J_{\mu\nu}(x)$ into the correlation functions
$\Pi(p)$, $\Pi_{\mu\nu}(p)$ and $\Pi_{\mu\nu\alpha\beta}(p)$ to obtain the hadronic representation
\cite{SVZ79-1,SVZ79-2,PRT85,QiaoCF-review-2025},  and isolate the  lowest pentaquark  states with negative and positive parities,
\begin{eqnarray}\label{CF-Hadron-12}
\Pi(p) & = & {\lambda^{-}_{\frac{1}{2}}}^2  {\!\not\!{p}+ M_{-} \over M_{-}^{2}-p^{2}  }+  {\lambda^{+}_{\frac{1}{2}}}^2  {\!\not\!{p}- M_{+} \over M_{+}^{2}-p^{2}  } +\cdots  \, ,\nonumber\\
&=&\Pi_{\frac{1}{2}}^1(p^2)\!\not\!{p}+\Pi_{\frac{1}{2}}^0(p^2)\, ,
 \end{eqnarray}
\begin{eqnarray}\label{CF-Hadron-32}
 \Pi_{\mu\nu}(p) & = & {\lambda^{-}_{\frac{3}{2}}}^2  {\!\not\!{p}+ M_{-} \over M_{-}^{2}-p^{2}  } \left(- g_{\mu\nu}+\frac{\gamma_\mu\gamma_\nu}{3}+\frac{2p_\mu p_\nu}{3p^2}-\frac{p_\mu\gamma_\nu-p_\nu \gamma_\mu}{3\sqrt{p^2}}
\right)\nonumber\\
&&+  {\lambda^{+}_{\frac{3}{2}}}^2  {\!\not\!{p}- M_{+} \over M_{+}^{2}-p^{2}  } \left(- g_{\mu\nu}+\frac{\gamma_\mu\gamma_\nu}{3}+\frac{2p_\mu p_\nu}{3p^2}-\frac{p_\mu\gamma_\nu-p_\nu \gamma_\mu}{3\sqrt{p^2}}
\right)   \nonumber \\
& &+ {f^{+}_{\frac{1}{2}}}^2  {\!\not\!{p}+ M_{+} \over M_{+}^{2}-p^{2}  } p_\mu p_\nu+  {f^{-}_{\frac{1}{2}}}^2  {\!\not\!{p}- M_{-} \over M_{-}^{2}-p^{2}  } p_\mu p_\nu  +\cdots  \, , \nonumber\\
&=&\left[\Pi_{\frac{3}{2}}^1(p^2)\!\not\!{p}+\Pi_{\frac{3}{2}}^0(p^2)\right]\left(- g_{\mu\nu}\right)+\cdots\, ,
\end{eqnarray}
\begin{eqnarray}\label{CF-Hadron-52}
\Pi_{\mu\nu\alpha\beta}(p) & = &2{\lambda^{-}_{\frac{5}{2}}}^2  {\!\not\!{p}+ M_{-} \over M_{-}^{2}-p^{2}  } \left[\frac{ \widetilde{g}_{\mu\alpha}\widetilde{g}_{\nu\beta}+\widetilde{g}_{\mu\beta}\widetilde{g}_{\nu\alpha}}{2}-\frac{\widetilde{g}_{\mu\nu}\widetilde{g}_{\alpha\beta}}{5}-\frac{1}{10}\left( \gamma_{\mu}\gamma_{\alpha}+\frac{\gamma_{\mu}p_{\alpha}-\gamma_{\alpha}p_{\mu}}{\sqrt{p^2}}-\frac{p_{\mu}p_{\alpha}}{p^2}\right)\widetilde{g}_{\nu\beta}\right.\nonumber\\
&&\left.-\frac{1}{10}\left( \gamma_{\nu}\gamma_{\alpha}+\frac{\gamma_{\nu}p_{\alpha}-\gamma_{\alpha}p_{\nu}}{\sqrt{p^2}}-\frac{p_{\nu}p_{\alpha}}{p^2}\right)\widetilde{g}_{\mu\beta}
+\cdots\right]\nonumber\\
&&+  2 {\lambda^{+}_{\frac{5}{2}}}^2  {\!\not\!{p}- M_{+} \over M_{+}^{2}-p^{2}  } \left[\frac{ \widetilde{g}_{\mu\alpha}\widetilde{g}_{\nu\beta}+\widetilde{g}_{\mu\beta}\widetilde{g}_{\nu\alpha}}{2}
-\frac{\widetilde{g}_{\mu\nu}\widetilde{g}_{\alpha\beta}}{5}-\frac{1}{10}\left( \gamma_{\mu}\gamma_{\alpha}+\frac{\gamma_{\mu}p_{\alpha}-\gamma_{\alpha}p_{\mu}}{\sqrt{p^2}}-\frac{p_{\mu}p_{\alpha}}{p^2}\right)\widetilde{g}_{\nu\beta}\right.\nonumber\\
&&\left.
-\frac{1}{10}\left( \gamma_{\nu}\gamma_{\alpha}+\frac{\gamma_{\nu}p_{\alpha}-\gamma_{\alpha}p_{\nu}}{\sqrt{p^2}}-\frac{p_{\nu}p_{\alpha}}{p^2}\right)\widetilde{g}_{\mu\beta}
 +\cdots\right]   \nonumber\\
 && +{f^{+}_{\frac{3}{2}}}^2  {\!\not\!{p}+ M_{+} \over M_{+}^{2}-p^{2}  } \left[ p_\mu p_\alpha \left(- g_{\nu\beta}+\frac{\gamma_\nu\gamma_\beta}{3}+\frac{2p_\nu p_\beta}{3p^2}-\frac{p_\nu\gamma_\beta-p_\beta \gamma_\nu}{3\sqrt{p^2}}
\right)+\cdots \right]\nonumber\\
&&+  {f^{-}_{\frac{3}{2}}}^2  {\!\not\!{p}- M_{-} \over M_{-}^{2}-p^{2}  } \left[ p_\mu p_\alpha \left(- g_{\nu\beta}+\frac{\gamma_\nu\gamma_\beta}{3}+\frac{2p_\nu p_\beta}{3p^2}-\frac{p_\nu\gamma_\beta-p_\beta \gamma_\nu}{3\sqrt{p^2}}
\right)+\cdots \right]   \nonumber \\
& &+ {g^{-}_{\frac{1}{2}}}^2  {\!\not\!{p}+ M_{-} \over M_{-}^{2}-p^{2}  } p_\mu p_\nu p_\alpha p_\beta+  {g^{+}_{\frac{1}{2}}}^2  {\!\not\!{p}- M_{+} \over M_{+}^{2}-p^{2}  } p_\mu p_\nu p_\alpha p_\beta  +\cdots \, , \nonumber\\
& = & \left[\Pi_{\frac{5}{2}}^1(p^2)\!\not\!{p}+\Pi_{\frac{5}{2}}^0(p^2)\right]\left( g_{\mu\alpha}g_{\nu\beta}+g_{\mu\beta}g_{\nu\alpha}\right)  +\cdots \, ,
 \end{eqnarray}
where $\widetilde{g}_{\mu\nu}=g_{\mu\nu}-\frac{p_{\mu}p_{\nu}}{p^2}$. We adopt  the components $\Pi_{j}^{1/0}(p^2)$ with $j=\frac{1}{2}$, $\frac{3}{2}$, $\frac{5}{2}$ to exclude  any contaminations from other decuplet $qssc\bar{c}$ pentaquark states having  different spins.

Then we obtain the  spectral densities on the hadron (H) side through  dispersion relation,
\begin{eqnarray}\label{H-spectral-1}
\frac{{\rm Im}\Pi^1_j(s)}{\pi}&=& \lambda_{-}^2 \delta\left(s-M_{-}^2\right)+\lambda_{+}^2 \delta\left(s-M_{+}^2\right) =\, \rho^1_{H}(s) \, ,\\
\frac{{\rm Im}\Pi^0_j(s)}{\pi}&=&M_{-}\lambda_{-}^2 \delta\left(s-M_{-}^2\right)-M_{+}\lambda_{+}^2 \delta\left(s-M_{+}^2\right)
=\rho^0_{H}(s) \, ,
\end{eqnarray}
where $j=\frac{1}{2}$, $\frac{3}{2}$, $\frac{5}{2}$,
then we employ  weight functions $\sqrt{s}\exp\left(-\frac{s}{T^2}\right)$ and $\exp\left(-\frac{s}{T^2}\right)$ to achieve the expressions on the hadron side,
\begin{eqnarray}
\int_{4m_c^2}^{s_0}ds \left[\sqrt{s}\,\rho^1_{H}(s)+\rho^0_{H}(s)\right]\exp\left( -\frac{s}{T^2}\right)
&=&2M_{-}\lambda_{-}^2\exp\left( -\frac{M_{-}^2}{T^2}\right) \, ,
\end{eqnarray}
\begin{eqnarray}
\int_{4m_c^2}^{s^\prime_0}ds \left[\sqrt{s}\,\rho^1_{H}(s)-\rho^0_{H}(s)\right]\exp\left( -\frac{s}{T^2}\right)
&=&2M_{+}\lambda_{+}^2\exp\left( -\frac{M_{+}^2}{T^2}\right) \, ,
\end{eqnarray}
where the $s_0$ and $s_0^\prime$ are the continuum thresholds  for the decuplet $qssc\bar{c}$ pentaquark states to exclude contaminations from higher resonances and continuum states,  the $T^2$ is de facto Borel parameter. In this way, we  separate   contributions  of the decuplet $qssc\bar{c}$  pentaquark states having  negative and positive parities plainly   \cite{Wang1508-EPJC,WangHuang-EPJC-1508-12,WangZG-EPJC-1509-12,WangZG-NPB-1512-32,
WangZhang-APPB}.

At this time,  we carry out  the OPE with the help of the full $u$, $d$, $s$ and $c$ quark propagators, then obtain  the spectral densities on the QCD side through   dispersion relation,
\begin{eqnarray}\label{QCD-rho}
 \rho^1_{QCD}(s) &=&\frac{{\rm Im}\Pi^1_j(s)}{\pi}\, , \nonumber\\
\rho^0_{QCD}(s) &=&\frac{{\rm Im}\Pi^0_j(s)}{\pi}\, ,
\end{eqnarray}
where $j=\frac{1}{2}$, $\frac{3}{2}$, $\frac{5}{2}$. As an example, we present  explicit expressions of the QCD spectral densities for the current $J^1(x)$ in  Appendix,  and neglect  length expressions of other spectral densities for simplicity.
 We calculate  the quark-gluon operators up to dimension $13$ and order $\mathcal{O}( \alpha_s^{k})$ with $k\leq 1$ systematically  to achieve   the vacuum expectations, and   the terms  $\propto m_s$ to symbolize   the light-flavor   $SU(3)$ breaking effects \cite{WangZG-Review,WangZG-IJMPA-3-scheme,WangZG-Pc12-JpsiLambda,WangZG-Pc12-Jpsip,WZG-penta-IJMPA,
 WangZG-Pc12-JpsiXi,WangZG-Pc12-JpsiSgm,WangZG-Pcs4459-333}. In calculations, we resort to vacuum saturation to factorize the higher dimensional vacuum condensates into lower dimensional ones, the factorization hypothesis works well in the large-$N_c$ limit \cite{SVZ79-1,SVZ79-2}.

At last,  we  equal the QCD side to the hadron side of the selected components of the correlation functions $\Pi_{\frac{1}{2}/\frac{3}{2}/\frac{5}{2}}^{1/0}(p^2)$, accomplish the quark-hadron duality and employ  weight functions $\sqrt{s}\exp\left(-\frac{s}{T^2}\right)$ and $\exp\left(-\frac{s}{T^2}\right)$ again,  and  achieve  two   elegant   QCD sum rules,
\begin{eqnarray}\label{QCDSR}
2M_{-}\lambda_{-}^2\exp\left( -\frac{M_{-}^2}{T^2}\right)&=& \int_{4m_c^2}^{s_0}ds \,\left[\sqrt{s}\rho_{QCD}^1(s)+\rho_{QCD}^{0}(s)\right]\,\exp\left( -\frac{s}{T^2}\right)\,  ,
\end{eqnarray}
\begin{eqnarray}\label{QCDSR-Positive}
2M_{+}\lambda_{+}^2\exp\left( -\frac{M_{+}^2}{T^2}\right)&=& \int_{4m_c^2}^{s^\prime_0}ds \,\left[\sqrt{s}\rho_{QCD}^1(s)-\rho_{QCD}^{0}(s)\right]\,\exp\left( -\frac{s}{T^2}\right)\,  .
\end{eqnarray}
As we focus on the  decuplet $qssc\bar{c}$ pentaquark states with negative parity, it is enough to  retain only the QCD sum rules in  Eq.\eqref{QCDSR}. We derive    Eq.\eqref{QCDSR} in regard    to  $\frac{1}{T^2}$, then delete  the pole residues $\lambda_{-}$ to achieve  concise  QCD sum rules for the  pentaquark masses,
 \begin{eqnarray}\label{QCDSR-mass}
 M^2_{-} &=& \frac{-\int_{4m_c^2}^{s_0}ds \frac{d}{d(1/T^2)}\, \left[\sqrt{s}\rho_{QCD}^1(s)+\rho_{QCD}^{0}(s)\right]\,\exp\left( -\frac{s}{T^2}\right)}{\int_{4m_c^2}^{s_0}ds \, \left[\sqrt{s}\rho_{QCD}^1(s)+\rho_{QCD}^{0}(s)\right]\,\exp\left( -\frac{s}{T^2}\right)}\,  ,
\end{eqnarray}
thereafter, we would like to use $M_P$ ($\lambda_P$) in stead of $M_{-}$ ($\lambda_{-}$) to denote the masses (pole residues) of the hidden-charm pentaquark states with the negative parity.

\section{Numerical results and discussions}
Initially, we employ   conventional values of the  condensates as usual,
$\langle\bar{q}q \rangle=-(0.24\pm 0.01\, \rm{GeV})^3$,  $\langle\bar{s}s \rangle=(0.8\pm0.1)\langle\bar{q}q \rangle$,
 $\langle\bar{q}g_s\sigma G q \rangle=m_0^2\langle \bar{q}q \rangle$, $\langle\bar{s}g_s\sigma G s \rangle=m_0^2\langle \bar{s}s \rangle$,
$m_0^2=(0.8 \pm 0.1)\,\rm{GeV}^2$, $\langle \frac{\alpha_s
GG}{\pi}\rangle=0.012\pm0.004\,\rm{GeV}^4$    at the characteristic  energy scale  $\mu=1\, \rm{GeV}$
\cite{SVZ79-1,SVZ79-2,PRT85,ColangeloReview}, and employ  the $\overline{MS}$  masses $m_{c}(m_c)=(1.275\pm0.025)\,\rm{GeV}$
 and $m_s(\mu=2\,\rm{GeV})=(0.095\pm0.005)\,\rm{GeV}$
 listed in  The Review of Particle Physics \cite{PDG}.
In addition,  we take  the energy-scale dependence into account  \cite{Narison-mix},
 \begin{eqnarray}
 \langle\bar{q}q \rangle(\mu)&=&\langle\bar{q}q\rangle({\rm 1 GeV})\left[\frac{\alpha_{s}({\rm 1 GeV})}{\alpha_{s}(\mu)}\right]^{\frac{12}{33-2n_f}}\, , \nonumber\\
 \langle\bar{s}s \rangle(\mu)&=&\langle\bar{s}s \rangle({\rm 1 GeV})\left[\frac{\alpha_{s}({\rm 1 GeV})}{\alpha_{s}(\mu)}\right]^{\frac{12}{33-2n_f}}\, , \nonumber\\
 \langle\bar{q}g_s \sigma Gq \rangle(\mu)&=&\langle\bar{q}g_s \sigma Gq \rangle({\rm 1 GeV})\left[\frac{\alpha_{s}({\rm 1 GeV})}{\alpha_{s}(\mu)}\right]^{\frac{2}{33-2n_f}}\, ,\nonumber\\
  \langle\bar{s}g_s \sigma Gs \rangle(\mu)&=&\langle\bar{s}g_s \sigma Gs \rangle({\rm 1 GeV})\left[\frac{\alpha_{s}({\rm 1 GeV})}{\alpha_{s}(\mu)}\right]^{\frac{2}{33-2n_f}}\, ,\nonumber\\
m_c(\mu)&=&m_c(m_c)\left[\frac{\alpha_{s}(\mu)}{\alpha_{s}(m_c)}\right]^{\frac{12}{33-2n_f}} \, ,\nonumber\\
m_s(\mu)&=&m_s({\rm 2GeV} )\left[\frac{\alpha_{s}(\mu)}{\alpha_{s}({\rm 2GeV})}\right]^{\frac{12}{33-2n_f}}\, ,\nonumber\\
\alpha_s(\mu)&=&\frac{1}{b_0t}\left[1-\frac{b_1}{b_0^2}\frac{\log t}{t} +\frac{b_1^2(\log^2{t}-\log{t}-1)+b_0b_2}{b_0^4t^2}\right]\, ,
\end{eqnarray}
  where $t=\log \frac{\mu^2}{\Lambda^2}$, $b_0=\frac{33-2n_f}{12\pi}$, $b_1=\frac{153-19n_f}{24\pi^2}$, $b_2=\frac{2857-\frac{5033}{9}n_f+\frac{325}{27}n_f^2}{128\pi^3}$,  $\Lambda_{QCD}=210\,\rm{MeV}$, $292\,\rm{MeV}$  and  $332\,\rm{MeV}$ for the flavors  $n_f=5$, $4$ and $3$, respectively  \cite{PDG}.

As we study  the $qssc\bar{c}$  pentaquark states,  we  employ the flavor numbers $n_f=4$, then evolve all the input parameters to a characteristic energy scale $\mu$ obeying the modified energy scale formula,
\begin{eqnarray}\label{Modify-formula}
\mu &=&\sqrt{M_{P}^2-(2{\mathbb{M}}_c)^2}-2{\mathbb{M}}_s \, ,
 \end{eqnarray}
 with the effective quark masses ${\mathbb{M}}_c$ and ${\mathbb{M}}_s$,  which generally and collectively epitomize the heavy degrees of freedom and light-flavor $SU(3)$ breaking effects, respectively, the best fitted values are ${\mathbb{M}}_c=1.82\,\rm{GeV}$ and ${\mathbb{M}}_s=0.15\,\rm{GeV}$ \cite{WangZG-Review,WangZG-Pc12-JpsiLambda,WangZG-Pc12-Jpsip,
WangZG-Pc12-JpsiXi,WangZG-Pc12-JpsiSgm,WangHuang3900, Wang-tetra-formula,WangZG-mole-formula-1,WangZG-mole-formula-2,
Wang-tetra-NPA-2014}.

In the heavy quark limit, the heavy quark $Q$ serves as a  static well potential and  attracts  a light quark  to form a heavy diquark  in  color antitriplet. The heavy antiquark $\overline{Q}$ serves as another  static well potential and  attracts  a light diquark  to form a heavy  triquark  in  color triplet. Then the diquark and  triquark attract each other to form a  pentaquark state.

Thus, the hidden-heavy  pentaquark states are characterized by the effective $Q$-quark mass ${\mathbb{M}}_Q$ and the virtuality $V=\sqrt{M_{P}^2-(2{\mathbb{M}}_Q)^2}$ or the energy scale $\mu=\sqrt{M_{P}^2-(2{\mathbb{M}}_Q)^2}$ of the QCD spectral densities \cite{Wang1508-EPJC,WangHuang-EPJC-1508-12,WangZG-EPJC-1509-12,
WangZG-NPB-1512-32}. We set $m_u=m_d=0$, then introduce an effective $s$-quark mass ${\mathbb{M}}_s$ to epitomize the light-flavor $SU(3)$ breaking effects collectively.

We rewrite the modified energy scale formula in the form,
\begin{eqnarray}\label{formula-Regge}
M^2_{P}&=&(\mu+\delta)^2+{\rm Constant}\, ,
\end{eqnarray}
where the Constant has the value $4{\mathbb{M}}_c^2$ and $\delta=2{\mathbb{M}}_s$.  The predicted pentaquark masses and  pertinent energy scales of the QCD spectral densities have a
 Regge-trajectory-like relation.

The (modified) energy scale formula can strengthen the pole contributions significantly    and ameliorate the convergent behavior of the OPE  significantly \cite{WangZG-Review,WangZG-IJMPA-3-scheme}.
As in previous works,  we  constrain   the continuum threshold parameters as $\sqrt{s_0}=M_{P}+ (0.5-0.8)\,\rm{GeV}$
\cite{WangZG-Review,WangZG-Pc12-JpsiLambda,WangZG-Pc12-Jpsip,
WangZG-Pc12-JpsiXi,WangZG-Pc12-JpsiSgm}.

 At the beginning, we choose the energy scale $\mu=1.0\,\rm{GeV}$ tentatively, then optimize the continuum threshold parameters and Borel parameters to obtain the  masses  $M_{P}$, if they cannot satisfy the relation $\mu =\sqrt{M_{P}^2-(2{\mathbb{M}}_c)^2}-2{\mathbb{M}}_s$, then we set the energy scale $\mu=1.1\,\rm{GeV}$, $1.2\,\rm{GeV}$, $1.3\,\rm{GeV}$, $\cdots$ tentatively, and repeat
the same  procedure until obtaining  self-consistent  results.

After repeatedly trial  and error, we achieve the  Borel  windows and continuum thresholds, which are presented  plainly in Table \ref{Borel}. The pole contributions are about $(40-60)\%$, the largest pole contributions achieved  up to now in literature. We  certainly fulfill the standard of  pole dominance, and
  we define the pole contributions as usual,
\begin{eqnarray}
{\rm{pole}}&=&\frac{\int_{4m_{c}^{2}}^{s_{0}}ds\,\rho_{QCD}\left(s\right)\exp\left(-\frac{s}{T^{2}}\right)} {\int_{4m_{c}^{2}}^{\infty}ds\,\rho_{QCD}\left(s\right)\exp\left(-\frac{s}{T^{2}}\right)}\, ,
\end{eqnarray}
 with $\rho_{QCD}=\sqrt{s}\rho_{QCD}^1(s)+\rho_{QCD}^{0}(s)$.

\begin{figure}
\centering
\includegraphics[totalheight=8cm,width=10cm]{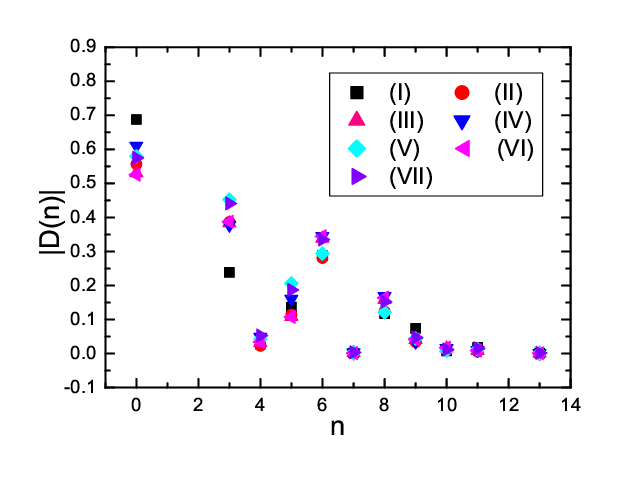}
  \caption{ The $|D(n)|$ via variations of $n$ in the case of central values of the input parameters, where the (I), (II), (III), (IV), (V), (VI)  and (VII)  denote the
   $[ss][qc]\bar{c}+2[sq][sc]\bar{c}$ ($1$, $1$, $0$, $\frac{1}{2}$),
$[ss][qc]\bar{c}+2[sq][sc]\bar{c}$ ($1$, $0$, $1$, $\frac{1}{2}$),
$[ss][qc]\bar{c}+2[sq][sc]\bar{c}$ ($1$, $0$, $1$, $\frac{3}{2}$),
$[ss][qc]\bar{c}+2[sq][sc]\bar{c}$ ($1$, $1$, $2$, $\frac{3}{2}$)${}_2$,
$[ss][qc]\bar{c}+2[sq][sc]\bar{c}$ ($1$, $1$, $2$, $\frac{3}{2}$)${}_3$,
$[ss][qc]\bar{c}+2[sq][sc]\bar{c}$ ($1$, $0$, $1$, $\frac{5}{2}$) and
$[ss][qc]\bar{c}+2[sq][sc]\bar{c}$ ($1$, $1$, $2$, $\frac{5}{2}$)  pentaquark states, respectively. }\label{OPE-fig}
\end{figure}

 In Fig.\ref{OPE-fig}, we plot contributions of the vacuum condensates with  dimension $n$ under the condition of  central values of  other  parameters, again we define  the $D(n)$ as usual,
   \begin{eqnarray}
D(n)&=&\frac{\int_{4m_{c}^{2}}^{s_{0}}ds\,\rho_{QCD,n}(s)\exp\left(-\frac{s}{T^{2}}\right)}
{\int_{4m_{c}^{2}}^{s_{0}}ds\,\rho_{QCD}\left(s\right)\exp\left(-\frac{s}{T^{2}}\right)}\, .
\end{eqnarray}
The largest contribution comes from the perturbative term, which characterizes the directional behavior of the OPE,   the $D(4)$ and $D(7)$ are small enough and neglectful, while the $D(6)$ plays  an   unique role and serves as a non-substitutable  milestone. The contributions $|D(n)|$  with $n\geq 6$ have obvious hierarchies,
\begin{eqnarray}
&&D(6)\gg |D(8)| \gg D(9) \gg D(10)\sim|D(11)| \gg D(13) \, ,
\end{eqnarray}
 the convergent behaviors of the OPE are very well.

 In Refs.\cite{Penta-NLO-Groote-PRD-2006,Penta-NLO-Groote-PRD-2012},  Groote,  Korner and Pivovarov  calculated the next-to-leading  order (NLO)  QCD corrections to perturbative contributions for the correlation functions of the five-quark currents  in the limit of massless quarks, and obtained very large NLO corrections. In the present case, the  largest contributions come from the perturbative terms, if the NLO corrections are large, just like in the case of massless quarks, the perturbative contributions would dominate the QCD sum rules, it is difficult to obtain stable Borel platforms. We have to calculate the NLO corrections to the quark condensates at least, if the radiative corrections are taken into account consistently, the predictions would be  improved   remarkably. The calculations are
 formidably  complex in the case of massive quarks, up to now, there lack literatures on this aspect. We would like to begin this work this year but expect to accomplish it in several years.

\begin{table}
\begin{center}
\begin{tabular}{|c|c|c|c|c|c|c|c|}\hline\hline
                  &$T^2(\rm{GeV}^2)$     &$\sqrt{s_0}(\rm{GeV})$    &$\mu(\rm{GeV})$  &pole          &$D(13)$         \\ \hline

$J^1(x)$          &$3.5-3.9$             &$5.39\pm0.10$             &$2.7$            &$(40-60)\%$   &$\ll 1\%$      \\ \hline

$J^2(x)$          &$3.8-4.2$             &$5.50\pm0.10$             &$2.8$            &$(42-60)\%$   &$\ll 1\%$       \\ \hline

$J^1_\mu(x)$      &$3.8-4.2$             &$5.47\pm0.10$             &$2.8$            &$(42-61)\%$   &$\ll 1\%$     \\ \hline

$J^2_\mu(x)$      &$3.9-4.3$             &$5.56\pm0.10$             &$2.9$            &$(41-60)\%$   &$\ll1\%$     \\ \hline

$J^3_\mu(x)$      &$3.7-4.1$             &$5.47\pm0.10$             &$2.8$            &$(42-61)\%$   &$\ll1\%$     \\ \hline

$J^1_{\mu\nu}(x)$ &$3.8-4.2$             &$5.44\pm0.10$             &$2.8$            &$(41-60)\%$   &$\ll1\%$     \\ \hline

$J^2_{\mu\nu}(x)$ &$3.8-4.2$             &$5.50\pm0.10$             &$2.8$            &$(42-61)\%$   &$\ll1\%$     \\ \hline
\hline
\end{tabular}
\end{center}
\caption{ The Borel  windows, continuum thresholds, optimal  energy scales, pole contributions,   contributions of  $D(13)$  for the decuplet  $qssc\bar{c}$ pentaquark states with  negative parity. }\label{Borel}
\end{table}

\begin{table}
\begin{center}
\begin{tabular}{|c|c|c|c|c|c|c|c|c|}\hline\hline
$[qq][qc]\bar{c}$ ($S_L$, $S_H$, $J_{LH}$, $J$) &$M(\rm{GeV})$   &$\lambda(10^{-3}\rm{GeV}^6)$ &\cite{WangZG-EPJC-1509-12,WangZG-NPB-1512-32}      \\ \hline

$[ss][qc]\bar{c}+2[sq][sc]\bar{c}$ ($1$, $1$, $0$, $\frac{1}{2}$)  &$4.71\pm0.12$ &$6.28\pm1.07$  &$4.58\pm0.14$              \\ \hline

$[ss][qc]\bar{c}+2[sq][sc]\bar{c}$ ($1$, $0$, $0$, $\frac{1}{2}$)  &$4.80\pm0.10$ &$7.63\pm1.15$  &$4.60\pm 0.11 $              \\ \hline

$[ss][qc]\bar{c}+2[sq][sc]\bar{c}$ ($1$, $0$, $1$, $\frac{3}{2}$)  &$4.77\pm0.10$ &$3.99\pm0.59$  &$4.60\pm 0.11 $              \\ \hline

$[ss][qc]\bar{c}+2[sq][sc]\bar{c}$ ($1$, $1$, $2$, $\frac{3}{2}$)${}_2$ &$4.87\pm0.10$  &$7.86\pm1.14$  &$4.62 \pm 0.11$ \\ \hline

$[ss][qc]\bar{c}+2[sq][sc]\bar{c}$ ($1$, $1$, $2$, $\frac{3}{2}$)${}_3$ &$4.78\pm0.10$   &$6.88\pm1.04$ &$4.61\pm 0.12$    \\ \hline

$[ss][qc]\bar{c}+2[sq][sc]\bar{c}$ ($1$, $0$, $1$, $\frac{5}{2}$)  &$4.75\pm0.10$ &$3.86\pm0.57$  &                   \\ \hline

$[ss][qc]\bar{c}+2[sq][sc]\bar{c}$ ($1$, $1$, $2$, $\frac{5}{2}$)  &$4.81\pm0.10$   &$3.91\pm0.57$  &                \\ \hline\hline
\end{tabular}
\end{center}
\caption{ The masses  and pole residues of the decuplet  $qssc\bar{c}$ pentaquark states with negative parity. In the last column, we present the predictions of the masses in our previous works, where the unit is GeV.  }\label{mass-Pcs}
\end{table}

\begin{figure}
\centering
\includegraphics[totalheight=5cm,width=7cm]{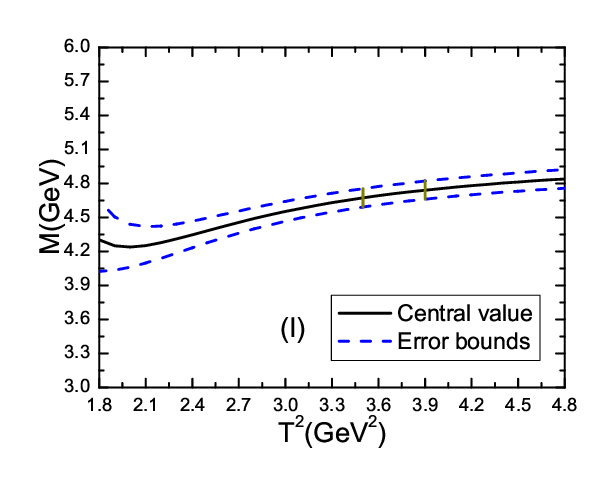}
\includegraphics[totalheight=5cm,width=7cm]{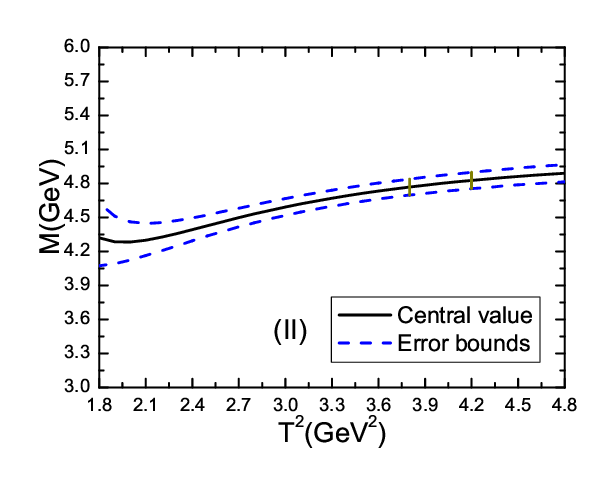}
\includegraphics[totalheight=5cm,width=7cm]{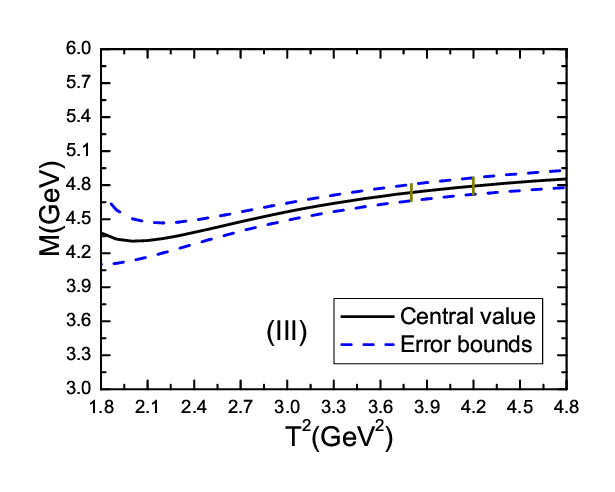}
\includegraphics[totalheight=5cm,width=7cm]{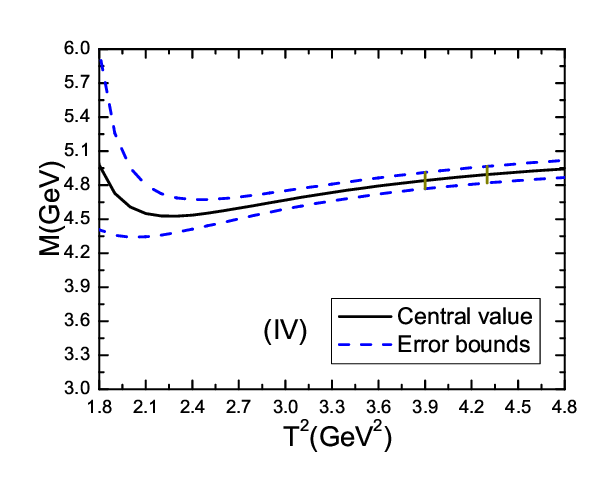}
\includegraphics[totalheight=5cm,width=7cm]{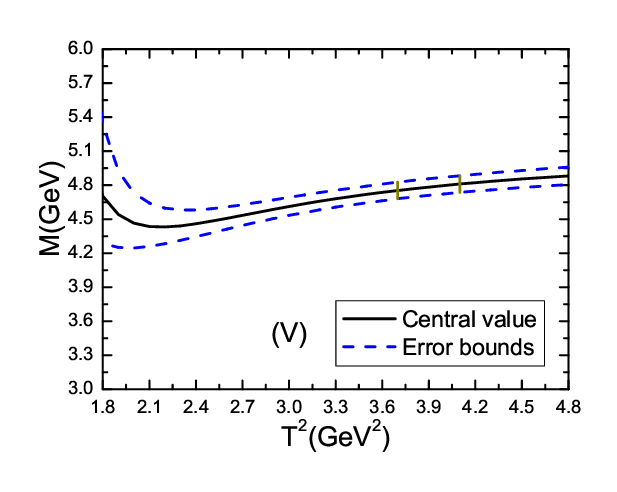}
\includegraphics[totalheight=5cm,width=7cm]{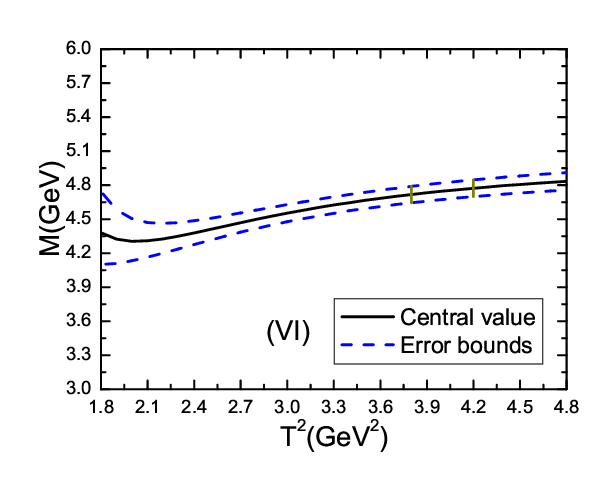}
\includegraphics[totalheight=5cm,width=7cm]{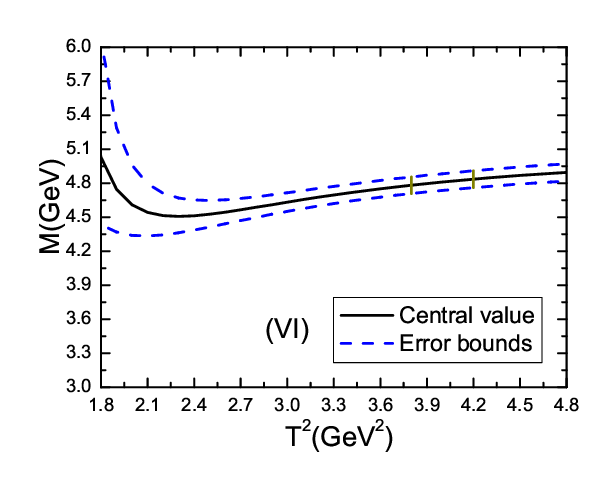}
  \caption{ The masses  via variations of the  Borel parameters $T^2$ for  the decuplet  $qssc\bar{c}$  pentaquark states, where the (I), (II), (III), (IV), (V),   (VI) and (VII)  denote the
   $[ss][qc]\bar{c}+2[sq][sc]\bar{c}$ ($1$, $1$, $0$, $\frac{1}{2}$),
$[ss][qc]\bar{c}+2[sq][sc]\bar{c}$ ($1$, $0$, $1$, $\frac{1}{2}$),
$[ss][qc]\bar{c}+2[sq][sc]\bar{c}$ ($1$, $0$, $1$, $\frac{3}{2}$),
$[ss][qc]\bar{c}+2[sq][sc]\bar{c}$ ($1$, $1$, $2$, $\frac{3}{2}$)${}_2$,
$[ss][qc]\bar{c}+2[sq][sc]\bar{c}$ ($1$, $1$, $2$, $\frac{3}{2}$)${}_3$,
$[ss][qc]\bar{c}+2[sq][sc]\bar{c}$ ($1$, $0$, $1$, $\frac{5}{2}$) and
  $[ss][qc]\bar{c}+2[sq][sc]\bar{c}$ ($1$, $1$, $2$, $\frac{5}{2}$) pentaquark states, respectively. }\label{mass-1-fig}
\end{figure}

Finally, we take  into account  all uncertainties  of the input   parameters,
and achieve  the masses and pole residues of
 the decuplet  $qssc\bar{c}$  pentaquark states with negative parity, which are illustrated and presented   plainly  in Figs.\ref{mass-1-fig} and Table \ref{mass-Pcs}. From Tables \ref{Borel}-\ref{mass-Pcs}, we  observe   that predicted masses satisfy the modified energy scale formula certainly. In Table \ref{mass-Pcs}, we also present the predictions of the pentaquark masses in our old calculations \cite{WangZG-EPJC-1509-12,WangZG-NPB-1512-32}, in which we took into account the vacuum condensates up to dimension 10 and neglected the gluon condensate, furthermore, we neglected the collective light-flavor $SU(3)$ breaking effects  $\mathbb{M}_s$ in the energy scale formula, it is not surprise that we obtain quite different predictions.

In Fig.\ref{mass-1-fig}, we plot the masses of the decuplet  $qssc\bar{c}$ pentaquark states with negative parity according to variations of the Borel parameters in a large range to show importance of the Borel windows, which  are indicted by two short vertical  lines. In the Borel windows, there really exist flat platforms, the uncertainties come from the Borel parameters are rather small.

\begin{figure}
\centering
\includegraphics[totalheight=6cm,width=7cm]{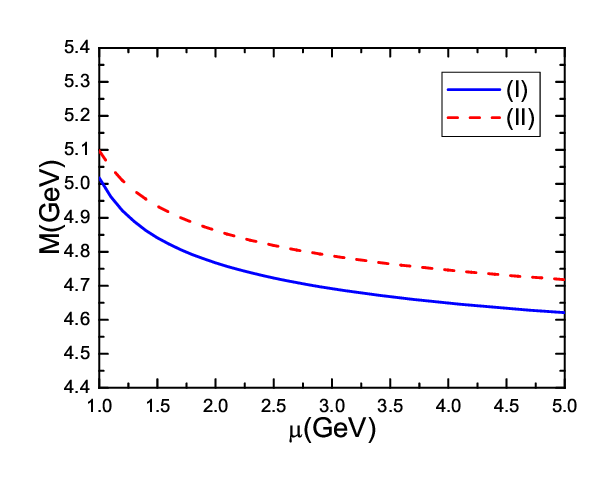}
\includegraphics[totalheight=6cm,width=7cm]{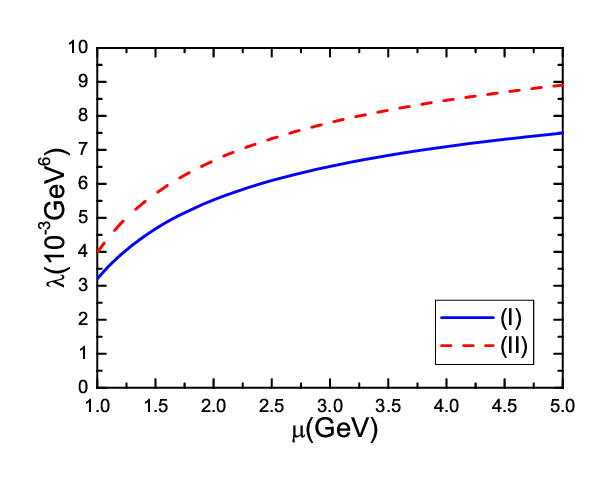}
  \caption{ The masses and pole residues  via variations of the  energy scales $\mu$ for  the decuplet  $qssc\bar{c}$  pentaquark states, where the (I) and (II) denote the
   $[ss][qc]\bar{c}+2[sq][sc]\bar{c}$ ($1$, $1$, $0$, $\frac{1}{2}$) and
$[ss][qc]\bar{c}+2[sq][sc]\bar{c}$ ($1$, $0$, $1$, $\frac{1}{2}$) pentaquark states, respectively. }\label{mass-mu-fig}
\end{figure}

In Fig.\ref{mass-mu-fig}, as an example, we plot the predicted masses and pole residues  via variations of the  energy scales $\mu$ for  the $[ss][qc]\bar{c}+2[sq][sc]\bar{c}$ ($1$, $1$, $0$, $\frac{1}{2}$) and
$[ss][qc]\bar{c}+2[sq][sc]\bar{c}$ ($1$, $0$, $1$, $\frac{1}{2}$) pentaquark states for the central values of other parameters. From the figure, we can see clearly that the predicted masses (pole residues) decrease (increase) monotonously with increase of the energy scales $\mu$, we need some constraints to determine the suitable energy scales of the QCD spectral densities and  adopt the modified  energy scale formula, see Eq.\eqref{Modify-formula}.

In the present work, we adopt vacuum saturation for the higher dimensional vacuum condensates,  derivations of the factorization hypothesis could be parameterized  by a parameters $\kappa$. We take account of non-factorizable contributions by taking the simple replacements,
\begin{eqnarray}
\langle \bar{q}q\rangle^2 &\to & \kappa\, \langle \bar{q}q\rangle^2 \, , \nonumber\\
\langle \bar{q}q\rangle \langle \frac{\alpha_sGG}{\pi}\rangle &\to &\kappa \,\langle \bar{q}q\rangle \langle \frac{\alpha_sGG}{\pi}\rangle\, , \nonumber\\
\langle \bar{q}q\rangle \langle \bar{q}g_s \sigma Gq\rangle &\to & \kappa\, \langle \bar{q}q\rangle\langle \bar{q}g_s \sigma Gq\rangle \, ,
\end{eqnarray}
etc, where $q=u$, $d$ or $s$.

\begin{figure}
\centering
\includegraphics[totalheight=6cm,width=7cm]{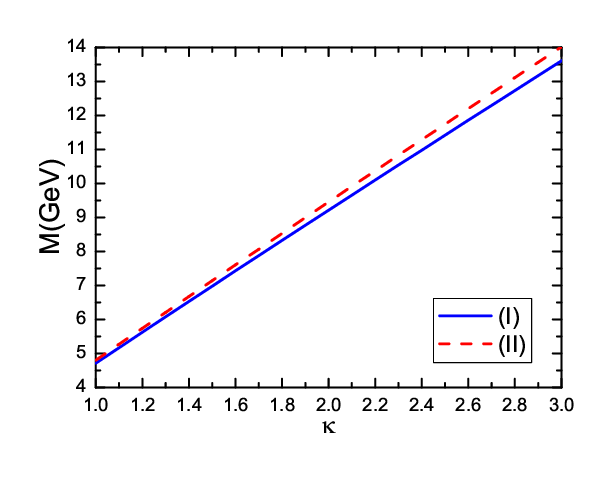}
  \caption{ The masses  via variations of the  parameter $\kappa$ for  the decuplet  $qssc\bar{c}$  pentaquark states, where the (I) and (II) denote the
   $[ss][qc]\bar{c}+2[sq][sc]\bar{c}$ ($1$, $1$, $0$, $\frac{1}{2}$) and
$[ss][qc]\bar{c}+2[sq][sc]\bar{c}$ ($1$, $0$, $1$, $\frac{1}{2}$) pentaquark states, respectively. }\label{mass-kapa-fig}
\end{figure}

In Fig.\ref{mass-kapa-fig}, again, as an example, we plot the predicted masses  via variations of the  parameter  $\kappa$ for  the $[ss][qc]\bar{c}+2[sq][sc]\bar{c}$ ($1$, $1$, $0$, $\frac{1}{2}$) and
$[ss][qc]\bar{c}+2[sq][sc]\bar{c}$ ($1$, $0$, $1$, $\frac{1}{2}$) pentaquark states for the central values of other parameters. From the figure, we can see clearly that the predicted masses increase monotonously and quickly  with increase of the parameter $\kappa$, the lowest values correspond to the factorization hypothesis, i.e $\kappa=1$. We usually take $\kappa=1$ to obtain satisfactory predictions compared to the experimental data on the multiquark candidates in the QCD sum rules \cite{WangZG-Review}.

In the present work, we take zero-width approximation for the hadronic spectral densities.  The interpolating currents  also couple potentially to the meson-baryon scattering states if they have the same quantum numbers, thus the  intermediate two-particle loops  contribute  a finite imaginary part to modify the dispersion relation on the hadron side \cite{WangZG-Review,WangZG-Landau-PRD}.
We can take  account of the finite width effect by the  simple replacement,
\begin{eqnarray}
\delta \left(s-M^2_{P} \right) &\to& \frac{1}{\pi}\frac{M_{P}\Gamma_{P}}{(s-M_{P}^2)^2+M_{P}^2\Gamma_{P}^2}\, .
\end{eqnarray}
Then the hadron  sides of  the QCD sum rules in Eq.\eqref{QCDSR} and Eq.\eqref{QCDSR-mass} undergo the  changes,
\begin{eqnarray}
\lambda^2_{P}\exp \left(-\frac{M^2_{P}}{T^2} \right) &\to& \lambda^2_{P}\int_{\Delta^2}^{s_0}ds\frac{1}{\pi}\frac{M_{P}\Gamma_{P}}{(s-M_{P}^2)^2+M_{P}^2\Gamma_{P}^2}\exp \left(-\frac{s}{T^2} \right)\, , \nonumber\\
&=&(0.87\sim0.88)\,\lambda^2_{P}\exp \left(-\frac{M^2_{P}}{T^2} \right)\, , \\
\lambda^2_{P}M^2_{P}\exp \left(-\frac{M^2_{P}}{T^2} \right) &\to& \lambda^2_{P}\int_{\Delta^2}^{s_0}ds\,s\,\frac{1}{\pi}\frac{M_{P}\Gamma_{P}}{
(s-M_{P}^2)^2+M_{P}^2\Gamma_{P}^2}\exp \left(-\frac{s}{T^2} \right)\, , \nonumber\\
&=&(0.87\sim0.87)\,\lambda^2_{P}M^2_{P}\exp \left(-\frac{M^2_{P}}{T^2} \right)\, ,
\end{eqnarray}
with the threshold $\Delta^2=(m_{\eta_c}+m_{\Xi^*})^2$ and continuum threshold  $\sqrt{s_0}=5.39\,\rm{GeV}$  by setting the width of the
$[ss][qc]\bar{c}+2[sq][sc]\bar{c}$ ($1$, $1$, $0$, $\frac{1}{2}$) pentaquark state to be the typical value $\Gamma_P=100\,\rm{MeV}$.
We can absorb the numerical factors  $0.87\sim0.88$ and $0.87\sim0.87$ into the pole residue with the simple replacement $\lambda_{P}\to 0.93\sim 94\,\lambda_P$ safely. Thus  the predicted  pentaquark masses are almost unaffected  but   the pole residues are affected slightly.

We cannot identify a hadron un-ambiguously based on the mass alone. For further works, we intend to take the pole residues as basic  input parameters to investigate  the kinematically allowed two-body strong decays,
 \begin{eqnarray}
P_{css}&\to& \bar{D}_s\Xi^*_c(4547/4548)\, , \,\bar{D}^*_s\Xi^*_c(4690/4691)\, , \, \bar{D}\Omega_c(4560/4564)\, , \, \bar{D}^*\Omega_c(4702/4705)\, ,  \nonumber\\
&&\,J/\psi \Xi^*(4631/4629) \, , \, \eta_c \Xi^*(4518/4516) \, ,
\end{eqnarray}
 via the  QCD sum rules, and achieve  the partial decay widths and approximately total widths,  and choose the
best decay channels to search for the decuplet  $qssc\bar{c}$  states experimentally, where we add thresholds of the meson-baryon pairs in the bracket with  unit $\rm{MeV}$ from the Particle Data Group for  intuition and clarity  \cite{PDG}.
The $P_c(4312)$, $P_c(4337)$, $P_c(4380)$, $P_c(4440)$ and $P_c(4457)$  were observed in the $J/\psi p$ mass spectrum,  the $P_{cs}(4338)$ and $P_{cs}(4459)$ were observed in the $J/\psi \Lambda$  mass spectrum, we expect that other octet $P_{cs}$ and $P_{css}$ states could be observed in the $J/\psi \Sigma$ and $J/\psi \Xi$ mass spectra respectively \cite{WangZG-Pc12-JpsiXi,WangZG-Pc12-JpsiSgm}, and the decuplet $P$ states  could be observed  in the $J/\psi\Delta$, $J/\psi\Sigma^*$, $J/\psi\Xi^*$ $J/\psi\Omega$ mass spectra, respectively.

In the diquark-diquark-antiquark model for the pentaquark states, we classify the hidden-charm pentaquark states according to the light-flavor $SU(3)$ symmetry,
 \begin{eqnarray}
{\mathbf{3}}\otimes {\mathbf{3}}\otimes {\mathbf{3}} &\to &  {\mathbf{1}} \oplus {\mathbf{8}} \oplus {\mathbf{8}}\oplus {\mathbf{10}}\, ,
\end{eqnarray}
there exist  two light-flavor octets.
And we expect that the same argument survives for the light-flavor baryons, however, there exists only one octet for the ground states and the singlet is forbidden by Fermi-Dirac statistics. We usually classify the light-flavor baryons according to the $SU(6)$ symmetry, i.e. combine the light-flavor and spin together to form the $SU(6)$ symmetry,
\begin{eqnarray}
\mathbf{6} \otimes \mathbf{6} \otimes \mathbf{6} &\to& \mathbf{56} \oplus \mathbf{70} \oplus \mathbf{70}  \oplus \mathbf{20}\, .
\end{eqnarray}
 The $J^P={\frac{1}{2}}^+$ octet  and ${\frac{3}{2}}^+$ decuplet
make up the ground-state 56-plet,
for the two 70-plets and one 20-plet,   some excitations of the spatial part of the wave function have to be introduced in order to make
the overall wave function symmetric. As a byproduct,   we can re-testify the classifications of the light baryons by investigating  the pentaquark  decays, a hidden-charm pentaquark state prefer to decay to $J/\psi$ $+$ light baryon in the same $SU(3)$ representation as the pentaquark state.

\section{Conclusion}
 In this work, we adopt the method of the QCD sum rules and  explore the diquark-diquark-antiquark type  decuplet $qssc\bar{c}$ pentaquark  states extensively. We compute   the    vacuum condensates up to dimension $13$ consistently as we usually do, achieve  the spectral densities at the QCD side through dispersion relation, and distinguish  the contributions of the  decuplet $qssc\bar{c}$  states  with negative parity clearly by excluding contaminations  from the corresponding  states with positive parity.
 Then we pick up  the characteristic  energy scales of the QCD spectral densities according to  the modified energy scale formula considering the collective light-flavor $SU(3)$ breaking effects. In the end, we achieve  the spectroscopy of the lowest  decuplet $qssc\bar{c}$ pentaquark states with the quantum numbers $IJ^{P}=\frac{1}{2}{\frac{1}{2}}^-$, $\frac{1}{2}{\frac{3}{2}}^-$, $\frac{1}{2}{\frac{5}{2}}^-$. We suggest to search for the decuplet exotic states  experimentally in the future in the processes
$\Xi_b^{\prime0}
\to P_{css}^0\,\phi \to J/\psi \Xi^{*0}  \phi $ and
$\Omega_b^{-}\to P_{css}^-\, \bar{K}^0 \to J/\psi \Xi^{*-}\, \bar{K}^0$. As a byproduct, we can testify classifications of the light baryons by investigating  the pentaquark decays.

\section*{Appendix}
The explicit expressions of the QCD spectral densities for the current $J^1(x)$,
\begin{eqnarray}
\rho^{1/0}_{QCD}(s)&=&\sum_k \rho^{1/0}_{k}(s)\, ,
\end{eqnarray}
\begin{eqnarray}
\rho^{0}_{k}(s)&=&m_c\,\tilde{\rho}^{0}_{k}(s)\, ,
\end{eqnarray}
with $k=0$, $3$, $4$, $5$, $6$, $7$, $8$, $9$, $10$, $11$, $13$,
where
\begin{eqnarray}
\rho^{1}_{0}(s)&=& \frac{1}{61440\pi^8}\int dydz\, zy(1-y-z)^{4} (s-\bar{m}_{c}^{2})^{4}(8s-3\bar{m}_{c}^{2})\exp\left(-\frac{s}{T^{2}}\right)\nonumber\\
	&&+\frac{m_{s}m_{c}}{9216\pi^{8}} \int dy dz\, 	z(1-y-z)^{3}
		\left(s-\bar{m}_{c}^{2}\right)^{4}	\exp\left(-\frac{s}{T^{2}}\right) \, ,
\end{eqnarray}

\begin{eqnarray}
\rho^{1}_{3}(s)&=&-\frac{ m_{c}\big[\langle\bar{q}q\rangle+2\langle\bar{s}s\rangle\big]}{1152\pi^{6}}
		 \int dydz	\, z(1-y-z)^{2} (s-\bar{m}_{c}^{2})^{3}		
		\exp\left(-\frac{s}{T^{2}}\right) \nonumber\\	 &&+\frac{m_{s}\big[2\langle\bar{s}s\rangle-\langle\bar{q}q\rangle\big]}{192\pi^{6}}
		 \int dy dz	\,zy(1-y-z)^{2}\left(s-\bar{m}_{c}^{2}\right)^{2}
		\left(2s-\bar{m}_{c}^{2}\right)				 \exp\left(-\frac{s}{T^{2}}\right) \, ,
\end{eqnarray}

\begin{eqnarray}
\rho^{1}_{4}(s)&=&	-\frac{m_{c}^{2}\left\langle\tilde{\alpha} GG\right\rangle}{9216\pi^{6}}
 \int dydz\frac{z(1-y-z)^{4}}{y^{2}}\left(s-\bar{m}_{c}^{2}\right)	 \left(5s-3\bar{m}_{c}^{2}\right)\exp\left(-\frac{s}{T^{2}}\right) \nonumber  \\
&&-\frac{\left\langle\tilde{\alpha} GG\right\rangle}{2048\pi^{6}}
\int dydz\, zy(1-y-z)^{2}\left(s-\bar{m}_{c}^{2}\right)^{2} \left(2s-\bar{m}_{c}^{2}\right) \exp\left(-\frac{s}{T^{2}}\right) \nonumber  \\
&&+\frac{\left\langle\tilde{\alpha} GG\right\rangle}{3072\pi^{6}}
\int dydz\, z(1-y-z)^{3}\left(s-\bar{m}_{c}^{2}\right)^{2} \left(2s-\bar{m}_{c}^{2}\right)	\exp\left(-\frac{s}{T^{2}}\right) \nonumber  \\
&&-\frac{m_{s}m_{c}^{3}\left\langle\tilde{\alpha} GG\right\rangle}{6912\pi^{6}}
 \int dydz\left(1+\frac{z}{y}\right)\frac{(1-y-z)^{3}}{y^{2}}
\left(s-\bar{m}_{c}^{2}\right)\exp\left(-\frac{s}{T^{2}}\right) \nonumber  \\
&&+\frac{m_{s}m_{c}\left\langle\tilde{\alpha} GG\right\rangle}{4608\pi^{6}}
\int dydz\, \frac{z(1-y-z)^{3}}{y^{2}}\left(s-\bar{m}_{c}^{2}\right)^{2}
\exp\left(-\frac{s}{T^{2}}\right)  \nonumber  \\		
&&-\frac{7m_{s}m_{c}\left\langle\tilde{\alpha} GG\right\rangle}{6144\pi^{6}}
\int dydz\,z(1-y-z)\left(s-\bar{m}_{c}^{2}\right)^{2}
\exp\left(-\frac{s}{T^{2}}\right) \nonumber  \\
&&+\frac{m_{s}m_{c}\left\langle\tilde{\alpha} GG\right\rangle}{1536\pi^{6}}
\int dydz\, \frac{z(1-y-z)^{2}}{y}\left(s-\bar{m}_{c}^{2}\right)^{2}
\exp\left(-\frac{s}{T^{2}}\right)\, ,		
\end{eqnarray}

\begin{eqnarray}
\rho^{1}_{5}(s)&=&\frac{m_{c}\big[\langle\bar{q}Gq\rangle+2\langle\bar{s}Gs\rangle
\big]}{512\pi^{6}} \int dy dz	\, z(1-y-z) (s-\bar{m}_{c}^{2})^{2}	 \exp\left(-\frac{s}{T^{2}}\right) \nonumber\\  	
&&+\frac{m_{s}\big[\langle\bar{q}Gq\rangle-\langle\bar{s}Gs\rangle\big]}{384\pi^{6}}
\int dydz\,zy(1-y-z)\left(s-\bar{m}_{c}^{2}\right)
\left(5s-3\bar{m}_{c}^{2}\right)\exp\left(-\frac{s}{T^{2}}\right) \nonumber \\
&&+\frac{m_{s}\big[\langle\bar{s}Gs\rangle+\langle\bar{q}Gq\rangle\big]}{1536\pi^{6}}
\int dydz\, zy(1-y-z)\left(s-\bar{m}_{c}^{2}\right)		 \left(5s-3\bar{m}_{c}^{2}\right)\exp\left(-\frac{s}{T^{2}}\right) \nonumber  \\
&&-\frac{m_{s}\big[\langle\bar{s}Gs\rangle+\langle\bar{q}Gq\rangle\big]}{1536\pi^{6}}
\int dydz\, z(1-y-z)^{2}\left(s-\bar{m}_{c}^{2}\right) \left(5s-3\bar{m}_{c}^{2}\right)\exp\left(-\frac{s}{T^{2}}\right) \, ,	 \end{eqnarray}

\begin{eqnarray}
\rho^{1}_{6}(s)&=&\frac{\langle\bar{s}s\rangle^{2}
+2\langle\bar{q}q\rangle\langle\bar{s}s\rangle}{144\pi^{4}}
 \int dy dz\,  zy(1-y-z) (s-\bar{m}_{c}^{2})(5s-3\bar{m}_{c}^{2})		 \exp\left(-\frac{s}{T^{2}}\right) \nonumber\\
 &&-\frac{m_{s}m_{c}\big[\langle\bar{s}s\rangle^{2}
 -11\langle\bar{q}q\rangle\langle\bar{s}s\rangle\big]}{72\pi^{4}}
		 \int dydz \, z	\left(s-\bar{m}_{c}^{2}\right)
		\exp\left(-\frac{s}{T^{2}}\right) \, ,
\end{eqnarray}

\begin{eqnarray}
\rho^{1}_{7}(s)&=&\frac{m_{c}^{3}\big[2\langle\bar{s}s\rangle
+\langle\bar{q}q\rangle\big]
\left\langle\tilde{\alpha} GG\right\rangle}{3456\pi^{4}}
 \int dy dz	\left(1+\frac{z}{y}\right)\frac{(1-y-z)^{2}}{y^2}
\exp\left(-\frac{s}{T^{2}}\right) \nonumber  \\
&&-\frac{m_{c}\big[2\langle\bar{s}s\rangle+\langle\bar{q}q\rangle\big]
\left\langle\tilde{\alpha} GG\right\rangle}{1152\pi^{4}}
\int dydz \frac{z(1-y-z)^{2}}{y^{2}}\left(s-\bar{m}_{c}^{2}\right)
\exp\left(-\frac{s}{T^{2}}\right)  \nonumber  \\			
&&+\frac{17m_{c}\big[2\langle\bar{s}s\rangle+\langle\bar{q}q\rangle\big]
\left\langle\tilde{\alpha} GG\right\rangle}{13824\pi^{4}}
\int dydz\, z\left(s-\bar{m}_{c}^{2}\right)
\exp\left(-\frac{s}{T^{2}}\right) \nonumber  \\
&&-\frac{m_{c}\big[2\langle\bar{s}s\rangle+\langle\bar{q}q\rangle\big]
\left\langle\tilde{\alpha} GG\right\rangle}{576\pi^{4}}
\int dydz\frac{z(1-y-z)}{y}	\left(s-\bar{m}_{c}^{2}\right)
\exp\left(-\frac{s}{T^{2}}\right) \nonumber  \\
&&-\frac{m_{s}m_{c}^{2}
\big[2\langle\bar{s}s\rangle-\langle\bar{q}q\rangle\big]
\left\langle\tilde{\alpha} GG\right\rangle}{288\pi^{4}}
 \int dydz	\frac{z(1-y-z)^{2}}{y^{2}} \left[1+\frac{s}{3}\delta\left(s-\bar{m}_{c}^{2}\right)\right]
\exp\left(-\frac{s}{T^{2}}\right) \nonumber  \\		
&&-\frac{m_{s}\langle\bar{s}s\rangle\left\langle\tilde{\alpha} GG\right\rangle}{1152\pi^{4}}\int dydz\, zy \left(4s-3\bar{m}_{c}^{2}\right)
\exp\left(-\frac{s}{T^{2}}\right) \nonumber  \\
&&+\frac{5m_{s}\langle\bar{s}s\rangle\left\langle\tilde{\alpha} GG\right\rangle}{1152\pi^{4}}\int dydz\, z(1-y-z)		 \left(4s-3\bar{m}_{c}^{2}\right)\exp\left(-\frac{s}{T^{2}}\right) \nonumber  \\
&&-\frac{m_{s}\big[\langle\bar{q}q\rangle-\langle\bar{s}s\rangle\big]
\left\langle\tilde{\alpha} GG\right\rangle}{768\pi^{4}}
\int dydz\, z(1-y-z) \left(4s-3\bar{m}_{c}^{2}\right)
\exp\left(-\frac{s}{T^{2}}\right) \nonumber  \\
&&-\frac{m_{s}\big[\langle\bar{s}s\rangle+\langle\bar{q}q\rangle\big]
\left\langle\tilde{\alpha} GG\right\rangle}{1728\pi^{4}}
\int dydz\, zy \left(4s-3\bar{m}_{c}^{2}\right)
\exp\left(-\frac{s}{T^{2}}\right) \, ,
\end{eqnarray}

\begin{eqnarray}
\rho^{1}_{8}(s)&=&-\frac{5\left[\langle\bar{s}s\rangle\langle\bar{s}Gs\rangle
+\langle\bar{s}s\rangle\langle\bar{q}Gq\rangle
+\langle\bar{q}q\rangle\langle\bar{s}Gs\rangle\right]}{576\pi^{4}}
\int dydz\, zy (4s-3\bar{m}_{c}^{2})	 \exp\left(-\frac{s}{T^{2}}\right)\nonumber\\
&&+\frac{\langle\bar{s}s\rangle\langle\bar{s}Gs\rangle
+\langle\bar{s}s\rangle\langle\bar{q}Gq\rangle
+\langle\bar{q}q\rangle\langle\bar{s}Gs\rangle}{288\pi^{4}}
\int dydz\, z(1-y-z)\left(4s-3\bar{m}_{c}^{2}\right)
\exp\left(-\frac{s}{T^{2}}\right) \nonumber  \\
&&+\frac{m_{s}m_{c}\big[5\langle\bar{s}s\rangle\langle\bar{s}Gs\rangle
-33\langle\bar{s}s\rangle\langle\bar{q}Gq\rangle-34\langle\bar{q}
q\rangle\langle\bar{s}Gs\rangle\big]}{864\pi^{4}}
 \int  dy \,(1-y)\exp\left(-\frac{s}{T^{2}}\right) \nonumber\\		
&&+\frac{m_{s}m_{c}\big[\langle\bar{s}s\rangle\langle\bar{s}Gs\rangle
+\langle\bar{s}s\rangle\langle\bar{q}Gq\rangle}{1152\pi^{4}}
\int dy\, (1-y)	\exp\left(-\frac{s}{T^{2}}\right) \nonumber  \\
&&+\frac{m_{s}m_{c}\big[\langle\bar{s}s\rangle\langle\bar{q}Gq\rangle
+\langle\bar{q}q\rangle\langle\bar{s}Gs\rangle}{72\pi^{4}}
\int dydz \frac{z}{y}\exp\left(-\frac{s}{T^{2}}\right) \, ,
\end{eqnarray}

\begin{eqnarray}
\rho^{1}_{9}(s)&=&-\frac{m_{c}\langle\bar{q}q\rangle\langle\bar{s}s\rangle^{2}}{9\pi^{2}}
	 \int dy (1-y) 	\exp\left(-\frac{s}{T^{2}}\right)\nonumber\\	
&&+\frac{m_{s}\langle\bar{q}q\rangle\langle\bar{s}s\rangle^{2}}{18\pi^{2}}
		 \int dy\, y(1-y)		 \left[1+\frac{s}{3}\delta\left(s-\tilde{m}_{c}^{2}\right)\right]
		\exp\left(-\frac{s}{T^{2}}\right)\, ,
\end{eqnarray}

\begin{eqnarray}
\rho^{1}_{10}(s)&=&\frac{\langle\bar{s}Gs\rangle^{2}
+2\langle\bar{q}Gq\rangle\langle\bar{s}Gs\rangle}{256\pi^{4}}
\int dy \, y(1-y) 	\left[1+\frac{s}{3}\delta(s-\tilde{m}_{c}^{2})
\right]	\exp\left(-\frac{s}{T^{2}}\right) \nonumber\\		
&&-\frac{m_{c}^{2}\big[\langle\bar{s}s\rangle^{2}+2\langle\bar{q}q\rangle\langle\bar{s}s
\rangle\big]\left\langle\tilde{\alpha} GG\right\rangle}{324\pi^{2}}
 \int dy dz	\frac{z(1-y-z)}{y^{2}} \left(1+\frac{s}{2T^{2}}\right)
\delta\left(s-\bar{m}_{c}^{2}\right)\exp\left(-\frac{s}{T^{2}}\right) \nonumber  \\
&&-\frac{\big[\langle\bar{s}s\rangle^{2}+2\langle\bar{q}q\rangle\langle\bar{s}s\rangle\big]
\left\langle\tilde{\alpha} GG\right\rangle}{1152\pi^{2}}
\int dydz\, z \left[1+\frac{s}{3}\delta\left(s-\bar{m}_{c}^{2}\right)\right]
\exp\left(-\frac{s}{T^{2}}\right) \nonumber  \\
&&-\frac{5\left[\langle\bar{s}Gs\rangle^{2}
+2\langle\bar{q}Gq\rangle\langle\bar{s}Gs\rangle\right]}{1536\pi^{4}}
 \int dydz\, z \left[1+\frac{s}{3}\delta\left(s-\bar{m}_{c}^{2}\right)\right]
\exp\left(-\frac{s}{T^{2}}\right) \nonumber  \\
&&+\frac{\big[\langle\bar{s}s\rangle^{2}+2\langle\bar{q}q\rangle\langle\bar{s}s\rangle\big]
\left\langle\tilde{\alpha} GG\right\rangle}{432\pi^{2}}
\int dy\,y(1-y) \left[1+\frac{s}{3}\delta\left(s-\tilde{m}_{c}^{2}\right)\right]
\exp\left(-\frac{s}{T^{2}}\right) \nonumber  \\	
&&-\frac{m_{s}m_{c}\big[\langle\bar{s}Gs\rangle^{2}
-17\langle\bar{q}Gq\rangle\langle\bar{s}Gs\rangle\big]}{864\pi^{4}}
\int dy\,(1-y)\left(1+\frac{s}{2T^{2}}\right)
\delta\left(s-\tilde{m}_{c}^{2}\right)\exp\left(-\frac{s}{T^{2}}\right) \nonumber\\			 &&+\frac{m_{s}m_{c}^{3}\big[\langle\bar{s}s\rangle^{2}
-11\langle\bar{q}q\rangle\langle\bar{s}s\rangle\big]\left\langle\tilde{\alpha} GG\right\rangle}{1296\pi^{2}T^{2}}\int dydz \frac{1}{y^{2}}\left(1+\frac{z}{y}\right)\delta\left(s-\bar{m}_{c}^{2}\right) \exp\left(-\frac{s}{T^{2}}\right) \nonumber  \\
&&-\frac{m_{s}m_{c}\big[\langle\bar{s}s\rangle^{2}
-11\langle\bar{q}q\rangle\langle\bar{s}s\rangle\big]
\left\langle\tilde{\alpha} GG\right\rangle}{432\pi^{2}}
\int dydz\frac{z}{y^{2}} \delta\left(s-\bar{m}_{c}^{2}\right) \exp\left(-\frac{s}{T^{2}}\right)\nonumber  \\
&&-\frac{m_{s}m_{c}\big[\langle\bar{s}s\rangle^{2}
+\langle\bar{q}q\rangle\langle\bar{s}s\rangle\big]
\left\langle\tilde{\alpha} GG\right\rangle}{864\pi^{2}}
 \int dy\frac{(1-y)}{y}	\delta\left(s-\tilde{m}_{c}^{2}\right) \exp\left(-\frac{s}{T^{2}}\right) \nonumber  \\
&&-\frac{m_{s}m_{c}\big[\langle\bar{s}Gs\rangle^{2}
+\langle\bar{s}Gs\rangle\langle\bar{q}Gq\rangle\big]}{3456\pi^{4}}
\int dy \, (1-y) \left(1+\frac{s}{2T^{2}}\right)
\delta\left(s-\tilde{m}_{c}^{2}\right) \exp\left(-\frac{s}{T^{2}}\right) \nonumber  \\	
&&-\frac{m_{s}m_{c}\langle\bar{q}Gq\rangle\langle\bar{s}Gs\rangle}{144\pi^{4}}
\int dy \frac{(1-y)}{y}	\delta\left(s-\tilde{m}_{c}^{2}\right) \exp\left(-\frac{s}{T^{2}}\right) \nonumber  \\		
&&-\frac{m_{s}m_{c}\big[\langle\bar{s}s\rangle^{2}-23\langle\bar{q}q\rangle\langle\bar{s}s\rangle\big]
\left\langle\tilde{\alpha} GG\right\rangle}{1296\pi^{2}}
\int dy\, (1-y)	 \left(1+\frac{s}{2T^{2}}\right)
\delta\left(s-\tilde{m}_{c}^{2}\right) \exp\left(-\frac{s}{T^{2}}\right) \, ,
\end{eqnarray}

\begin{eqnarray}
\rho^{1}_{11}(s)&=&\frac{m_{c}\big[\langle\bar{s}s\rangle^{2}\langle\bar{q}Gq\rangle
+2\langle\bar{q}q\rangle\langle\bar{s}s\rangle\langle\bar{s}Gs\rangle\big]}{18\pi^{2}}
\int dy\, (1-y) \left(1+\frac{s}{2T^{2}}\right)\delta(s-\tilde{m}_{c}^{2})
\exp\left(-\frac{s}{T^{2}}\right) \nonumber \\	
&&-\frac{m_{c}\big[\langle\bar{s}s\rangle^{2}\langle\bar{q}Gq\rangle
+2\langle\bar{q}q\rangle\langle\bar{s}s\rangle\langle\bar{s}Gs\rangle\big]}{108\pi^{2}}
\int dy \frac{(1-y)}{y}	\delta\left(s-\tilde{m}_{c}^{2}\right) \exp\left(-\frac{s}{T^{2}}\right) \nonumber  \\	
&&-\frac{m_{s}\big[3\langle\bar{s}s\rangle^{2}\langle\bar{q}Gq\rangle
+5\langle\bar{q}q\rangle\langle\bar{s}s\rangle\langle\bar{s}Gs\rangle\big]}{108\pi^{2}}
\int dy\,y(1-y)\left(1+\frac{2s}{3T^{2}}+\frac{s^2}{6T^{4}}\right)
\delta\left(s-\tilde{m}_{c}^{2}\right)\exp\left(-\frac{s}{T^{2}}\right) \nonumber \\	
&&+\frac{m_{s}\big[\langle\bar{s}s\rangle^{2}\langle\bar{q}Gq\rangle
+\langle\bar{q}q\rangle\langle\bar{s}s\rangle\langle\bar{s}Gs\rangle\big]}{216\pi^{2}}
\int dy\,(1-y) \left(1+\frac{s}{2T^{2}}\right) \delta\left(s-\tilde{m}_{c}^{2}\right)\exp\left(-\frac{s}{T^{2}}\right) \, ,
\end{eqnarray}

\begin{eqnarray}
\rho^{1}_{13}(s)&=&-\frac{m_{c}\big[\langle\bar{q}q\rangle\langle\bar{s}Gs\rangle^{2}
+2\langle\bar{s}s\rangle\langle\bar{s}Gs\rangle\langle\bar{q}Gq\rangle\big]}{72\pi^2T^2}
\int dy \, (1-y)\left(	1+\frac{s}{T^{2}}+\frac{s^2}{2T^{4}}\right)
\delta\left(s-\tilde{m}_{c}^{2}\right)\exp\left(-\frac{s}{T^{2}}\right) \nonumber \\
&&+\frac{m_{c}\big[\langle\bar{q}q\rangle\langle\bar{s}Gs\rangle^{2}
+2\langle\bar{s}s\rangle\langle\bar{q}Gq\rangle\langle\bar{s}Gs\rangle\big]}{216\pi^{2}T^{2}}
\int dy\, \frac{(1-y)}{y} \left(1+\frac{s}{T^{2}}\right)
\delta\left(s-\tilde{m}_{c}^{2}\right)\exp\left(-\frac{s}{T^{2}}\right) \nonumber  \\
&&+\frac{m_{c}^{3}\langle\bar{q}q\rangle\langle\bar{s}s\rangle^{2}
\left\langle\tilde{\alpha} GG\right\rangle}{162T^{4}} \int dy \frac{1}{y^3}		 \delta\left(s-\tilde{m}_{c}^{2}\right)\exp\left(-\frac{s}{T^{2}}\right) \nonumber  \\
&&-\frac{m_{c}\langle\bar{q}q\rangle\langle\bar{s}s\rangle^{2}
\left\langle\tilde{\alpha} GG\right\rangle}{54T^{2}}
\int dy\,\frac{(1-y)}{y^{2}}\delta\left(s-\tilde{m}_{c}^{2}\right) \exp\left(-\frac{s}{T^{2}}\right) \nonumber\\
&&-\frac{m_{c}\langle\bar{q}q\rangle\langle\bar{s}s\rangle^{2}
\left\langle\tilde{\alpha} GG\right\rangle}{54T^{2}}
\int dy \, (1-y) \left(1+\frac{s}{T^{2}}+\frac{s^2}{2T^{4}}\right)
\delta\left(s-\tilde{m}_{c}^{2}\right)\exp\left(-\frac{s}{T^{2}}\right) \nonumber \\
&&+\frac{m_{s}\big[2\langle\bar{q}q\rangle\langle\bar{s}Gs\rangle^{2}
+5\langle\bar{s}s\rangle\langle\bar{s}Gs\rangle\langle\bar{q}Gq\rangle\big]}{432\pi^{2}T^{2}}
 \int dy\, y(1-y)	 \left(1+\frac{s}{T^{2}}+\frac{s^2}{2T^4}+\frac{s^3}{6T^6}\right)		 \delta\left(s-\tilde{m}_{c}^{2}\right)\exp\left(-\frac{s}{M^{2}}\right)\nonumber\\
&&-\frac{m_{s}m_{c}^{2}\langle\bar{q}q\rangle\langle\bar{s}s\rangle^{2}
\left\langle\tilde{\alpha} GG\right\rangle}{486T^{6}}
\int dy\frac{(1-y)}{y^{2}}\, s\, \delta\left(s-\tilde{m}_{c}^{2}\right) \exp\left(-\frac{s}{M^{2}}\right) \nonumber\\
&&-\frac{m_{s}\big[\langle\bar{q}q\rangle\langle\bar{s}Gs\rangle^{2}
+4\langle\bar{s}s\rangle\langle\bar{q}Gq\rangle\langle\bar{s}Gs\rangle\big]}{1296\pi^{2}T^{2}}
\int dy \, (1-y) \left(1+\frac{s}{T^{2}}+\frac{s^2}{2T^{4}}\right)
\delta\left(s-\tilde{m}_{c}^{2}\right)\exp\left(-\frac{s}{T^{2}}\right) \nonumber \\
&&+\frac{m_{s}\langle\bar{q}q\rangle\langle\bar{s}s\rangle^{2}
\left\langle\tilde{\alpha} GG\right\rangle}{162T^{2}}
\int dy\, y(1-y) \left(1+\frac{s}{T^{2}}+\frac{s^{2}}{2T^{4}}+\frac{s^3}{6T^{6}}\right)
 \delta\left(s-\tilde{m}_{c}^{2}\right)\exp\left(-\frac{s}{T^{2}}\right)\, ,
\end{eqnarray}

\begin{eqnarray}
\tilde{\rho}_0^0(s)&=&\frac{1}{61440\pi^{8}} \int dydz
\, y(1-y-z)^{4} \left(s-\bar{m}_{c}^{2}\right)^{4}\left(7s-2\bar{m}_{c}^{2}\right)
\exp\left(-\frac{s}{T^{2}}\right) \nonumber \\
&&+\frac{m_{s}m_{c}}{9216\pi^{8}}\int dydz\, (1-y-z)^{3}	 \left(s-\bar{m}_{c}^{2}\right)^{4}\exp\left(-\frac{s}{T^{2}}\right) \, ,
\end{eqnarray}

\begin{eqnarray}
\tilde{\rho}_3^0(s)&=&-\frac{m_{c}\big[\langle\bar{q}q\rangle
+2\langle\bar{s}s\rangle\big]}{1152\pi^{6}}
\int dydz\, (1-y-z)^{2}\left(s-\bar{m}_{c}^{2}\right)^{3}
\exp\left(-\frac{s}{T^{2}}\right)  \nonumber \\
&&+\frac{m_{s}\big[2\langle\bar{s}s\rangle-\langle\bar{q}q\rangle\big]}{576\pi^{6}}
\int dydz\, y(1-y-z)^{2} \left(s-\bar{m}_{c}^{2}\right)^{2}\left(5s-2\bar{m}_{c}^{2}\right)			 \exp\left(-\frac{s}{T^{2}}\right) \, ,
\end{eqnarray}

\begin{eqnarray}
\tilde{\rho}_4^0(s)&=&
-\frac{m_{c}^{2}\left\langle\tilde{\alpha} GG\right\rangle}{9216\pi^{6}}
\int dydz \frac{(1-y-z)^{4}}{y^{2}}	\left(1+\frac{z}{y}\right)		 \left(s-\bar{m}_{c}^{2}\right)\left(2s-\bar{m}_{c}^{2}\right)
\exp\left(-\frac{s}{T^{2}}\right) \nonumber \\
&&+\frac{\left\langle\tilde{\alpha} GG\right\rangle}{18432\pi^{6}}
\int dydz\frac{z(1-y-z)^{4}}{y^{2}}\left(s-\bar{m}_{c}^{2}\right)^{2}		 \left(5s-2\bar{m}_{c}^{2}\right)\exp\left(-\frac{s}{T^{2}}\right) \nonumber \\	
&&-\frac{\left\langle\tilde{\alpha} GG\right\rangle}{6144\pi^{6}}
\int dydz\, y(1-y-z)^{2}\left(s-\bar{m}_{c}^{2}\right)^{2}	 \left(5s-2\bar{m}_{c}^{2}\right)\exp\left(-\frac{s}{T^{2}}\right) \nonumber \\
&&+\frac{\left\langle\tilde{\alpha} GG\right\rangle}{9216\pi^{6}}
\int dydz\,(1-y-z)^{3}\left(s-\bar{m}_{c}^{2}\right)^{2}		 \left(5s-2\bar{m}_{c}^{2}\right)\exp\left(-\frac{s}{T^{2}}\right) \nonumber \\
&&-\frac{m_{s}m_{c}^{3}\left\langle\tilde{\alpha} GG\right\rangle}{3456\pi^{6}}
\int dydz\frac{(1-y-z)^{3}}{y^{3}}\left(s-\bar{m}_{c}^{2}\right)
\exp\left(-\frac{s}{T^{2}}\right) \nonumber \\		
&&+\frac{m_{s}m_{c}\left\langle\tilde{\alpha} GG\right\rangle}{2304\pi^{6}}
\int dydz\frac{(1-y-z)^{3}}{y^{2}}\left(s-\bar{m}_{c}^{2}\right)^{2}
\exp\left(-\frac{s}{T^{2}}\right) \nonumber \\		
&&+\frac{m_{s}\big[2\langle\bar{s}s\rangle-\langle\bar{q}q\rangle\big]
\left\langle\tilde{\alpha} GG\right\rangle}{576\pi^{4}}
\int dydz\frac{z(1-y-z)^{2}}{y^{2}}	 \left(3s-2\bar{m}_{c}^{2}\right)
\exp\left(-\frac{s}{T^{2}}\right) \nonumber \\		
&&-\frac{7m_{s}m_{c}\left\langle\tilde{\alpha} GG\right\rangle}{6144\pi^{6}}
\int dydz\, (1-y-z)\left(s-\bar{m}_{c}^{2}\right)^{2}
\exp\left(-\frac{s}{T^{2}}\right)  \nonumber \\	
&&+\frac{m_{s}m_{c}\left\langle\tilde{\alpha} GG\right\rangle}{1536\pi^{6}}
\int dydz\,\frac{(1-y-z)^{2}}{y}\left(s-\bar{m}_{c}^{2}\right)^{2}
\exp\left(-\frac{s}{T^{2}}\right) \, ,			
\end{eqnarray}

\begin{eqnarray}
\tilde{\rho}_5^0(s)&=& \frac{m_{c}\big[\langle\bar{q}Gq\rangle+2\langle\bar{s}Gs\rangle\big]}{512\pi^{6}}
 \int dydz\,(1-y-z)\left(s-\bar{m}_{c}^{2}\right)^{2}		 \exp\left(-\frac{s}{T^{2}}\right)  \nonumber \\
&&+\frac{m_{s}\big[\langle\bar{q}Gq\rangle-\langle\bar{s}Gs\rangle\big]}{192\pi^{6}}
\int dy dz\, y(1-y-z)\left(s-\bar{m}_{c}^{2}\right)\left(2s-\bar{m}_{c}^{2}\right)
\exp\left(-\frac{s}{T^{2}}\right)  \nonumber \\		
&&+\frac{m_{s}\big[\langle\bar{s}Gs\rangle+\langle\bar{q}Gq\rangle\big]}{768\pi^{6}}
\int dydz\, y(1-y-z) \left(s-\bar{m}_{c}^{2}\right)\left(2s-\bar{m}_{c}^{2}\right) 		 \exp\left(-\frac{s}{T^{2}}\right) \nonumber \\ 			 &&-\frac{m_{s}\big[\langle\bar{q}Gq\rangle+\langle\bar{s}Gs\rangle\big]}{768\pi^{6}}
\int dydz\, (1-y-z)^{2}	\left(s-\bar{m}_{c}^{2}\right) \left(2s-\bar{m}_{c}^{2}\right)	\exp\left(-\frac{s}{T^{2}}\right) \, ,	 \end{eqnarray}

\begin{eqnarray}
\tilde{\rho}_6^0(s)&=& \frac{\langle\bar{s}s\rangle^{2}+2\langle\bar{q}q\rangle\langle\bar{s}s\rangle}{72\pi^{4}}
\int dydz\, y(1-y-z)\left(s-\bar{m}_{c}^{2}\right)\left(2s-\bar{m}_{c}^{2}\right)
\exp\left(-\frac{s}{T^{2}}\right)  \nonumber \\
&&-\frac{m_{s}m_{c}\big[\langle\bar{s}s\rangle^{2}
-11\langle\bar{q}q\rangle\langle\bar{s}s\rangle\big]}{72\pi^{4}}
\int dydz\left(s-\bar{m}_{c}^{2}\right)	\exp\left(-\frac{s}{T^{2}}\right) \, ,
\end{eqnarray}

\begin{eqnarray}
\tilde{\rho}_7^0(s)&=&  \frac{m_{c}^{3}\big[2\langle\bar{s}s\rangle+\langle\bar{q}q\rangle\big]
\left\langle\tilde{\alpha} GG\right\rangle}{1728\pi^{4}}
 \int dydz \frac{(1-y-z)^{2}}{y^{3}}\exp\left(-\frac{s}{T^{2}}\right)  \nonumber \\
&&-\frac{m_{c}\big[2\langle\bar{s}s\rangle+\langle\bar{q}q\rangle\big]
\left\langle\tilde{\alpha} GG\right\rangle}{576\pi^{4}}
 \int dydz\frac{(1-y-z)^{2}}{y^{2}}	\left(s-\bar{m}_{c}^{2}\right)
\exp\left(-\frac{s}{T^{2}}\right) \nonumber \\		
&&+\frac{17m_{c}\big[2\langle\bar{s}s\rangle+\langle\bar{q}q\rangle\big]
\left\langle\tilde{\alpha} GG\right\rangle}{13824\pi^{4}}
\int dydz\left(s-\bar{m}_{c}^{2}\right)\exp\left(-\frac{s}{T^{2}}\right) \nonumber \\
&&-\frac{m_{c}\big[2\langle\bar{s}s\rangle+\langle\bar{q}q\rangle\big]
\left\langle\tilde{\alpha} GG\right\rangle}{576\pi^{4}}
\int dydz\frac{(1-y-z)}{y}\left(s-\bar{m}_{c}^{2}\right)
\exp\left(-\frac{s}{T^{2}}\right) \nonumber \\	
&&-\frac{m_{s}m_{c}^{2}\big[2\langle\bar{s}s\rangle-\langle\bar{q}q\rangle\big]
\left\langle\tilde{\alpha} GG\right\rangle}{864\pi^{4}}
 \int dydz\frac{(1-y-z)^{2}}{y^{2}}	\left(1+\frac{z}{y}\right) \left[1+\frac{s}{2}\delta\left(s-\bar{m}_{c}^{2}\right)\right]
\exp\left(-\frac{s}{T^{2}}\right) \nonumber \\
&&-\frac{m_{s}\langle\bar{s}s\rangle
\left\langle\tilde{\alpha} GG\right\rangle}{1152\pi^{4}}
\int dydz\,y \left(3s-2\bar{m}_{c}^{2}\right)
\exp\left(-\frac{s}{T^{2}}\right) \nonumber \\
&&+\frac{5m_{s}\langle\bar{s}s\rangle
\left\langle\tilde{\alpha} GG\right\rangle}{1152\pi^{4}}
\int dydz\, (1-y-z) \left(3s-2\bar{m}_{c}^{2}\right)
\exp\left(-\frac{s}{T^{2}}\right) \nonumber \\
&&-\frac{m_{s}\big[\langle\bar{q}q\rangle-\langle\bar{s}s\rangle\big]
\left\langle\tilde{\alpha} GG\right\rangle}{768\pi^{4}}
\int dydz\,(1-y-z) \left(3s-2\bar{m}_{c}^{2}\right)
\exp\left(-\frac{s}{T^{2}}\right) \nonumber \\		
&&-\frac{m_{s}\big[\langle\bar{s}s\rangle+\langle\bar{q}q\rangle\big]
\left\langle\tilde{\alpha} GG\right\rangle}{1728\pi^{4}}
\int dydz\,y \left(3s-2\bar{m}_{c}^{2}\right)
\exp\left(-\frac{s}{T^{2}}\right) \, ,
\end{eqnarray}

\begin{eqnarray}
\tilde{\rho}_8^0(s)&=&-\frac{5\left[\langle\bar{q}q\rangle\langle\bar{s}Gs\rangle
+\langle\bar{s}s\rangle\langle\bar{s}Gs\rangle
+\langle\bar{s}s\rangle\langle\bar{q}Gq\rangle\right]}{576\pi^{4}}
\int dydz\,y \left(3s-2\bar{m}_{c}^{2}\right)
\exp\left(-\frac{s}{T^{2}}\right) \nonumber \\	
&&+\frac{\langle\bar{s}s\rangle\langle\bar{s}Gs\rangle
 +\langle\bar{s}s\rangle\langle\bar{q}Gq\rangle
 +\langle\bar{q}q\rangle\langle\bar{s}Gs\rangle}{288\pi^{4}}
 \int dy dz (1-y-z)	 \left(3s-2\bar{m}_{c}^{2}\right)
\exp\left(-\frac{s}{T^{2}}\right) \nonumber \\
&&+\frac{m_{s}m_{b}\big[5\langle\bar{s}s\rangle\langle\bar{s}Gs\rangle
-33\langle\bar{s}s\rangle\langle\bar{q}Gq\rangle
-34\langle\bar{q}q\rangle\langle\bar{s}Gs\rangle\big]}{864\pi^{4}}
\int dy		\exp\left(-\frac{s}{T^{2}}\right) \nonumber \\
&&+\frac{m_{s}m_{c}\big[\langle\bar{s}s\rangle\langle\bar{s}Gs\rangle
+\langle\bar{s}s\rangle\langle\bar{q}Gq\rangle\big]}{1152\pi^{4}}
\int dy	\exp\left(-\frac{s}{T^{2}}\right)\nonumber \\
&&+\frac{m_{s}m_{c}\big[\langle\bar{s}s\rangle\langle\bar{q}Gq\rangle
+\langle\bar{q}q\rangle\langle\bar{s}Gs\rangle\big]}{72\pi^{4}}
 \int dydz \frac{1}{y}	\exp\left(-\frac{s}{T^{2}}\right)\, ,
\end{eqnarray}

\begin{eqnarray}
\tilde{\rho}_9^0(s)&=&-\frac{m_{c}\langle\bar{q}q\rangle\langle\bar{s}s\rangle^{2}}{9\pi^{2}}
\int dy	\exp\left(-\frac{s}{T^{2}}\right)   \nonumber \\
&&+\frac{m_{s}\langle\bar{q}q\rangle\langle\bar{s}s\rangle^{2}}{27\pi^{2}}
\int dy\, y	\left[1+\frac{s}{2}\delta\left(s-\tilde{m}_{c}^{2}\right)	\right]
\exp\left(-\frac{s}{T^{2}}\right) \, ,
\end{eqnarray}

\begin{eqnarray}
\tilde{\rho}_{10}^0(s)&=& \frac{\langle\bar{s}Gs\rangle^{2}+2\langle\bar{q}Gq\rangle\langle\bar{s}Gs\rangle}{576\pi^{4}}
 \int dy\, y\left[1+\frac{s}{2}\delta\left(s-\tilde{m}_{c}^{2}\right)		\right]	 \exp\left(-\frac{s}{T^{2}}\right)  \nonumber \\
&&-\frac{m_{c}^{2}\big[\langle\bar{s}s\rangle^{2}
+2\langle\bar{q}q\rangle\langle\bar{s}s\rangle\big]
\left\langle\tilde{\alpha} GG\right\rangle}{1296\pi^{2}}
 \int dydz \frac{(1-y-z)}{y^{2}}\left(1+\frac{z}{y}\right)
\left(1+\frac{s}{T^{2}}\right)\delta\left(s-\bar{m}_{c}^{2}\right) \exp\left(-\frac{s}{T^{2}}\right) \nonumber \\
&&+\frac{\big[\langle\bar{s}s\rangle^{2}+2\langle\bar{q}q\rangle\langle\bar{s}s\rangle\big]
\left\langle\tilde{\alpha} GG\right\rangle}{216\pi^{2}}
 \int dydz	\frac{z(1-y-z)}{y^{2}} \left[1+\frac{s}{2}\delta\left(s-\bar{m}_{c}^{2}\right)\right]
\exp\left(-\frac{s}{T^{2}}\right) \nonumber \\ 		
&&-\frac{\big[\langle\bar{s}s\rangle^{2}+2\langle\bar{q}q\rangle\langle\bar{s}s\rangle\big]
\left\langle\tilde{\alpha} GG\right\rangle}{1728\pi^{2}}
 \int dydz \left[1+\frac{s}{2}\delta\left(s-\bar{m}_{c}^{2}\right)\right]
\exp\left(-\frac{s}{T^{2}}\right) \nonumber \\
&&+\frac{\langle\bar{s}Gs\rangle^{2}+2\langle\bar{s}Gs\rangle\langle\bar{q}Gq
\rangle}{1152\pi^{4}} \int dy\,y \left[1+\frac{s}{2}\delta\left(s-\tilde{m}_{c}^{2}\right)\right]
\exp\left(-\frac{s}{T^{2}}\right) \nonumber \\
&&-\frac{5\left[\langle\bar{s}Gs\rangle^{2}+2\langle\bar{q}Gq\rangle\langle\bar{s}Gs
\rangle\right]}{2304\pi^{4}} \int dydz \left[1+\frac{s}{2}\delta\left(s-\bar{m}_{c}^{2}\right)\right]
\exp\left(-\frac{s}{T^{2}}\right) \nonumber \\	
&&+\frac{\big[\langle\bar{s}s\rangle^{2}+2\langle\bar{q}q\rangle\langle\bar{s}s\rangle\big]
\left\langle\tilde{\alpha} GG\right\rangle}{648\pi^{2}}
\int dy\,y \left[1+\frac{s}{2}\delta\left(s-\tilde{m}_{c}^{2}\right)\right]
\exp\left(-\frac{s}{T^{2}}\right) \nonumber \\	
&&-\frac{m_{s}m_{c}\big[\langle\bar{s}Gs\rangle^{2}
-17\langle\bar{q}Gq\rangle\langle\bar{s}Gs\rangle\big]}{1728\pi^{4}}
\int dy	\left(1+\frac{s}{T^{2}}	\right)	\delta\left(s-\tilde{m}_{c}^{2}\right) \exp\left(-\frac{s}{T^{2}}\right) \nonumber \\		
&&+\frac{m_{s}m_{c}^{3}\big[\langle\bar{s}s\rangle^{2}
 -11\langle\bar{q}q\rangle\langle\bar{s}s\rangle\big]
 \left\langle\tilde{\alpha} GG\right\rangle}{648\pi^{2}T^{2}}
 \int dydz	\frac{1}{y^{3}}	\delta\left(s-\bar{m}_{c}^{2}\right) \exp\left(-\frac{s}{T^{2}}\right) \nonumber \\
&&-\frac{m_{s}m_{c}\big[\langle\bar{s}s\rangle^{2}
-11\langle\bar{q}q\rangle\langle\bar{s}s\rangle\big]
\left\langle\tilde{\alpha} GG\right\rangle}{216\pi^{2}}
 \int  dydz  \frac{1}{y^{2}}\delta\left(s-\bar{m}_{c}^{2}\right) \exp\left(-\frac{s}{T^{2}}\right) \nonumber \\ 			
&&-\frac{m_{s}m_{c}\big[\langle\bar{s}s\rangle^{2}
+\langle\bar{q}q\rangle\langle\bar{s}s\rangle\big]
\left\langle\tilde{\alpha} GG\right\rangle}{864\pi^{2}}
 \int dy \frac{1}{y}\delta\left(s-\tilde{m}_{c}^{2}\right) \exp\left(-\frac{s}{T^{2}}\right) \nonumber \\
&&-\frac{m_{s}m_{c}\big[\langle\bar{s}Gs\rangle^{2}
 +\langle\bar{s}Gs\rangle\langle\bar{q}Gq\rangle\big]}{6912\pi^{4}}
 \int dy \left(1+\frac{s}{T^{2}}\right)	\delta\left(s-\tilde{m}_{c}^{2}\right) \exp\left(-\frac{s}{T^{2}}\right) \nonumber \\	
&&-\frac{m_{s}m_{c}\langle\bar{q}Gq\rangle\langle\bar{s}Gs\rangle}{144\pi^{4}}
 \int dy\frac{1}{y}	\delta\left(s-\tilde{m}_{c}^{2}\right) \exp\left(-\frac{s}{T^{2}}\right) \nonumber \\
&&-\frac{m_{s}m_{c}\big[\langle\bar{s}s\rangle^{2}-23\langle\bar{q}q\rangle\langle\bar{s}s\rangle\big]
\left\langle\tilde{\alpha} GG\right\rangle}{2592\pi^{2}}
 \int dy \left(1+\frac{s}{T^{2}}\right)
\delta\left(s-\tilde{m}_{c}^{2}\right)\exp\left(-\frac{s}{T^{2}}\right) \, ,
\end{eqnarray}

\begin{eqnarray}
\tilde{\rho}_{11}^0(s)&=&\frac{m_{c}\big[\langle\bar{s}s\rangle^{2}\langle\bar{q}Gq\rangle
+2\langle\bar{q}q\rangle\langle\bar{s}s\rangle\langle\bar{s}Gs\rangle\big]}{36\pi^{2}}
 \int dy\left(1+\frac{s}{T^{2}}	\right)	\delta\left(s-\tilde{m}_{c}^{2}\right) \exp\left(-\frac{s}{T^{2}}\right)   \nonumber \\
&&-\frac{m_{c}\big[\langle\bar{s}s\rangle^{2}\langle\bar{q}Gq\rangle
+2\langle\bar{q}q\rangle\langle\bar{s}s\rangle\langle\bar{s}Gs\rangle\big]}{108\pi^{2}}
 \int dy	\frac{1}{y}	\delta\left(s-\tilde{m}_{c}^{2}\right) \exp\left(-\frac{s}{T^{2}}\right) \nonumber \\	
&&-\frac{m_{s}\big[3\langle\bar{s}s\rangle^{2}\langle\bar{q}Gq\rangle
+5\langle\bar{q}q\rangle\langle\bar{s}s\rangle\langle\bar{s}Gs\rangle\big]}{324\pi^{2}}
 \int_{y_{i}}^{y_{f}} dy\,y
\left(1+\frac{s}{T^{2}}+\frac{s^2}{2T^{4}}\right)		 \delta\left(s-\tilde{m}_{c}^{2}\right) \exp\left(-\frac{s}{T^{2}}\right) \nonumber \\		
&&+\frac{m_{s}\big[\langle\bar{s}s\rangle^{2}\langle\bar{q}Gq\rangle
 +\langle\bar{q}q\rangle\langle\bar{s}s\rangle\langle\bar{s}Gs\rangle\big]}{432\pi^{2}}
 \int dy \left(1+\frac{s}{T^{2}}\right)	\delta\left(s-\tilde{m}_{c}^{2}\right) \exp\left(-\frac{s}{M^{2}}\right) \, ,
\end{eqnarray}

\begin{eqnarray}
\tilde{\rho}_{13}^0(s)&=&-\frac{m_{c}\big[\langle\bar{q}q\rangle\langle\bar{s}Gs\rangle^{2}+2\langle\bar{s}s\rangle\langle\bar{s}Gs\rangle\langle\bar{q}Gq\rangle\big]
		}{144\pi^{2}T^{6}} \int dy\,s^2 \delta\left(s-\tilde{m}_{c}^{2}\right) \exp\left(-\frac{s}{T^{2}}\right) \nonumber \\
&&+\frac{m_{c}^{3}\langle\bar{q}q\rangle\langle\bar{s}s\rangle^{2}
\left\langle\tilde{\alpha} GG\right\rangle}{81T^{4}} \int dy\frac{1}{y^{3}}
\delta\left(s-\tilde{m}_{c}^{2}\right)\exp\left(-\frac{s}{T^{2}}\right) \nonumber \\
&&-\frac{m_{c}\langle\bar{q}q\rangle\langle\bar{s}s\rangle^{2}
\left\langle\tilde{\alpha} GG\right\rangle}{27T^{2}}
 \int dy\frac{1}{y^{2}} \delta\left(s-\tilde{m}_{c}^{2}\right) \exp\left(-\frac{s}{M^{2}}\right) \nonumber \\
&&+\frac{m_{c}\big[\langle\bar{q}q\rangle\langle\bar{s}Gs\rangle^{2}
+2\langle\bar{s}s\rangle\langle\bar{q}Gq\rangle\langle\bar{s}Gs\rangle\big]}{216\pi^{2}T^{4}}
 \int dy\frac{1}{y}	\,s\, \delta\left(s-\tilde{m}_{c}^{2}\right) \exp\left(-\frac{s}{M^{2}}\right) \nonumber \\
&&-\frac{m_{c}\langle\bar{q}q\rangle\langle\bar{s}s\rangle^{2}
\left\langle\tilde{\alpha} GG\right\rangle}{108T^{6}}
\int dy\, s^2\, \delta\left(s-\tilde{m}_{c}^{2}\right) \exp\left(-\frac{s}{T^{2}}\right) \nonumber \\		
&&+\frac{m_{s}\big[2\langle\bar{q}q\rangle\langle\bar{s}Gs\rangle^{2}
+5\langle\bar{s}s\rangle\langle\bar{s}Gs\rangle\langle\bar{q}Gq\rangle\big]}{2592\pi^{2}T^{8}}
 \int dy \,y \,s^3	\delta\left(s-\tilde{m}_{c}^{2}\right) \exp\left(-\frac{s}{M^{2}}\right)\nonumber \\
&&+\frac{m_{s}m_{c}^{2}\langle\bar{q}q\rangle\langle\bar{s}s\rangle^{2}
\left\langle\tilde{\alpha} GG\right\rangle}{972T^{4}}
 \int dy\frac{1}{y^{3}}	\left(1-\frac{s}{T^{2}}\right)
\delta\left(s-\tilde{m}_{c}^{2}\right)\exp\left(-\frac{s}{T^{2}}\right) \nonumber \\
&&+\frac{m_{s}\langle\bar{q}q\rangle\langle\bar{s}s\rangle^{2}
\left\langle\tilde{\alpha} GG\right\rangle}{324T^{4}} \int dy\frac{(1-y)}{y^{2}}	 s\,\delta\left(s-\tilde{m}_{c}^{2}\right) \exp\left(-\frac{s}{T^{2}}\right)	 \nonumber \\
&&-\frac{m_{s}\big[\langle\bar{q}q\rangle\langle\bar{s}Gs\rangle^{2}
 +4\langle\bar{s}s\rangle\langle\bar{q}Gq\rangle\langle\bar{s}Gs\rangle\big]	 }{2592\pi^{2}T^{6}} \int  dy\,s^2	\delta\left(s-\tilde{m}_{c}^{2}\right) \exp\left(-\frac{s}{M^{2}}\right) \nonumber \\
&&+\frac{m_{s}\langle\bar{q}q\rangle\langle\bar{s}s\rangle^{2}
\left\langle\tilde{\alpha} GG\right\rangle}{972T^{8}}
		 \int dy\,y	\,	s^{3}	 \delta\left(s-\tilde{m}_{c}^{2}\right)\exp\left(-\frac{s}{T^{2}}\right)\, ,
\end{eqnarray}
 $\langle\bar{q} Gq\rangle=\langle\bar{q}g_s\sigma Gq\rangle$, $q=u$, $d$, $s$, $\tilde{\alpha}_s=\frac{\alpha_s}{\pi}$,
$\int dydz=\int_{y_{i}}^{y_{f}} dy \int_{z_{i}}^{1-y} dz$,
$dy=\int_{y_{i}}^{y_{f}} dy $, $y_{f}=\frac{1+\sqrt{1-4m_c^2/s}}{2}$, $y_{i}=\frac{1-\sqrt{1-4m_c^2/s}}{2}$, $z_{i}=\frac{ym_c^2}{y s -m_c^2}$, $\bar{m}_c^2=\frac{(y+z)m_c^2}{yz}$,
$ \tilde{m}_c^2=\frac{m_c^2}{y(1-y)}$, and $\int_{y_i}^{y_f}dy \to \int_{0}^{1}dy$, $\int_{z_i}^{1-y}dz \to \int_{0}^{1-y}dz$ when the $\delta$ functions $\delta\left(s-\bar{m}_c^2\right)$ and $\delta\left(s-\tilde{m}_c^2\right)$ appear.
\section*{Acknowledgements}
This  work is supported by National Natural Science Foundation, Grant Number  12575083.


\begin{thebibliography}{99}

\bibitem{Gell-Mann-1964} M. Gell-Mann, Phys. Lett. {\bf 8} (1964) 214.

\bibitem{Strottman-1979} D. Strottman, Phys. Rev. {\bf D20} (1979) 748.

\bibitem{Lipkin-1987} H.  J. Lipkin, Phys. Lett. {\bf B195} (1987) 484.


\bibitem{exp2003-Cita1540-1}  T. Nakano {\it et al}, Phys. Rev. Lett. {\bf91} (2003) 012002.

\bibitem{exp2003-Cita1540-2} S. Stepanyan {\it et al}, Phys. Rev. Lett. {\bf
91} (2003) 252001.

\bibitem{exp2003-Cita1540-3} V. Kubarovsky {\it et al}, Phys. Rev. Lett. {\bf 92} (2004) 032001.

\bibitem{exp2004-cuudd}  A. Aktas {\it et al},  Phys. Lett. {\bf B588} (2004) 17.

\bibitem{LHCb-4380} R. Aaij  et al, Phys. Rev. Lett. {\bf 115} (2015) 072001.

\bibitem{LHCb-Pc4312} R. Aaij et al, Phys. Rev. Lett. {\bf 122} (2019) 222001.

\bibitem{Review-penta-mole-ZhSL-RPT}  H. X. Chen, W. Chen, X. Liu and S. L. Zhu,  Phys. Rept. {\bf 639} (2016) 1.

\bibitem{Review-penta-Esposito-RPT}  A. Esposito, A. Pilloni and A. D. Polosa, Phys. Rept. {\bf 668} (2017) 1.


\bibitem{Review-penta-Ali-PPNP} A. Ali, J. S. Lange and S. Stone, Prog. Part. Nucl. Phys. {\bf 97} (2017) 123.

\bibitem{Review-penta-mole-GuoFK-RMP}  F. K. Guo, C. Hanhart, U. G. Meissner, Q. Wang, Q. Zhao and B. S. Zou,  Rev. Mod. Phys. {\bf 90} (2018) 015004.


\bibitem{Review-penta-mole-LiuYR-PPNP}  Y. R. Liu, H. X. Chen, W. Chen, X. Liu and  S. L. Zhu, Prog. Part. Nucl. Phys. {\bf 107} (2019) 237.


\bibitem{Review-penta-mole-Brambilla-RPT} N. Brambilla, S. Eidelman, C. Hanhart, A. Nefediev, C. P. Shen, C. E. Thomas, A. Vairo and C. Z. Yuan,
Phys. Rept. {\bf 873} (2020) 1.


\bibitem{LHCb-Pcs4459-2012} R. Aaij  et al, Sci. Bull. {\bf 66} (2021) 1278.

\bibitem{LHCb-Pc4337}  R. Aaij  et al,  Phys. Rev. Lett. {\bf 128} (2022) 062001.

\bibitem{LHCb-Pcs4338}  R. Aaij et al, Phys. Rev. Lett. {\bf 131} (2023)  031901.

\bibitem{Belle-Pcs4338-Pcs4459} I. Adachi et al,  Phys. Rev. Lett. {\bf 135} (2025) 041901.


\bibitem{Review-mole-WangB-PRT}   L. Meng, B. Wang,  G. J. Wang and S. L. Zhu, Phys. Rept. {\bf 1019} (2023) 1.


\bibitem{Review-mole-GengLS-PRT} M. Z. Liu, Y. W. Pan, Z. W. Liu, T. W. Wu, J. X. Lu and L. S. Geng, Phys. Rept. {\bf 1108} (2025) 1.

\bibitem{WangZG-Review} Z. G. Wang,  Front. Phys. {\bf 21} (2026) 016300.


\bibitem{LHCb-LamcD-EPJC-2024} R. Aaij et al,  Eur. Phys. J. {\bf C84} (2024) 575.

\bibitem{LHCb-SigmaD-PRD-2024} R. Aaij et al, Phys. Rev. {\bf D110} (2024) L031104.

\bibitem{LHCb-LamcDs-PRD-2025} R. Aaij et al, Phys. Rev. {\bf D112} (2025) 052013.

\bibitem{LHCb-JpsiXi-EPJC-2025} R. Aaij et al, Eur. Phys. J. {\bf C85} (2025) 812.


\bibitem{AJBuras-1996} G. Buchalla, A. J. Buras and M. E. Lautenbacher,  Rev. Mod. Phys. {\bf 68} (1996) 1125.

\bibitem{di-di-anti-penta-1} L. Maiani, A. D. Polosa and V. Riquer,  Phys. Lett. {\bf B749} (2015) 289.

\bibitem{di-di-anti-penta-2} G. N. Li, M. He and X. G. He,  JHEP {\bf 1512} (2015) 128.

\bibitem{Oset-1} F. L. Wang, R. Chen and X. Liu, Phys. Rev. {\bf D103} (2021) 034014.

\bibitem{Oset-2}  L. Roca, J. Song and E. Oset, Phys. Rev. {\bf D109} (2024)  094005.

\bibitem{WangZG-IJMPA-3-scheme} Z. G. Wang, Int. J. Mod. Phys. {\bf A34} (2019)
    1950097.

\bibitem{Pcs4459-mole-WangZG-SR}  Z. G. Wang and Q. Xin, Chin. Phys. {\bf C45} (2021) 123105.


\bibitem{Pc4312-mole-penta-WXW-SCPMA}  X. W. Wang, Z. G. Wang, G. L. Yu and Q. Xin, Sci. China-Phys. Mech. Astron. {\bf 65} (2022) 291011.

\bibitem{Pc4312-mole-penta-WXW-IJMPA} X. W. Wang and Z. G. Wang, Int. J. Mod. Phys. {\bf A37} (2022) 2250189.


\bibitem{Pcs4338-mole-XWWang} X. W. Wang and Z. G. Wang, Chin. Phys. {\bf C47} (2023)  013109.

\bibitem{QCDSR-penta-mole-ChenHX-PRD-2019}   H. X. Chen,  W. Chen and S. L. Zhu,  Phys. Rev. {\bf D100} (2019)  051501.

\bibitem{QCDSR-penta-mole-ChenHX-EPJC-2021}   H. X. Chen, W. Chen, X. Liu and X. H. Liu,  Eur. Phys. J. {\bf C81} (2021) 409.

 \bibitem{QCDSR-penta-mole-Azizi-PRD-2017}  K. Azizi, Y. Sarac and H. Sundu,  Phys. Rev. {\bf D95} (2017)  094016.

 \bibitem{QCDSR-penta-mole-ZhangJR-EPJC-2019}  J. R. Zhang, Eur. Phys. J. {\bf C79} (2019)  1001.

 \bibitem{QCDSR-penta-mole-Azizi-PRD-2021} K. Azizi, Y. Sarac and H. Sundu, Phys. Rev. {\bf D103} (2021)  094033.



\bibitem{WangZG-Pc12-JpsiLambda}  Z. G. Wang and Q. Xin, Eur. Phys. J. {\bf C86} (2026)  472.

\bibitem{WangZG-Pc12-Jpsip}  Z. G. Wang, Eur. Phys. J. {\bf C86} (2026) 209.



\bibitem{Wang1508-EPJC} Z. G. Wang, Eur. Phys. J. {\bf C76} (2016) 70.

\bibitem{WangHuang-EPJC-1508-12} Z. G. Wang  and T. Huang, Eur. Phys. J. {\bf C76} (2016)  43.

\bibitem{WangZG-EPJC-1509-12} Z. G. Wang, Eur. Phys. J. {\bf C76} (2016)  142.

\bibitem{WangZG-NPB-1512-32} Z. G. Wang, Nucl. Phys. {\bf B913} (2016) 163.

\bibitem{WangZhang-APPB} J. X. Zhang, Z. G. Wang and Z. Y. Di,  Acta Phys. Polon. {\bf B48} (2017) 2013.


\bibitem{WZG-penta-IJMPA} Z. G. Wang, Int. J. Mod. Phys. {\bf A35} (2020)  2050003.


\bibitem{WangZG-Pc12-JpsiXi}  Z. G. Wang and Y. Liu, Nucl. Phys. {\bf B1027} (2026) 117456.

\bibitem{WangZG-Pc12-JpsiSgm}  Z. G. Wang and Y. Liu, arXiv: 2603.10774 [hep-ph].


\bibitem{SVZ79-1}  M. A. Shifman, A. I. Vainshtein and V. I. Zakharov, Nucl. Phys. {\bf B147} (1979) 385.

\bibitem{SVZ79-2}  M. A. Shifman, A. I. Vainshtein and V. I. Zakharov, Nucl. Phys. {\bf B147} (1979) 448.


\bibitem{PRT85} L. J. Reinders, H. Rubinstein and S. Yazaki, Phys. Rept. {\bf 127} (1985) 1.

\bibitem{QiaoCF-review-2025} S. Q. Zhang and C. F. Qiao,
arXiv: 2512.24706 [hep-ph].



\bibitem{WangZG-Pcs4459-333} Z. G. Wang, Int. J. Mod. Phys. {\bf A36} (2021) 2150071.

\bibitem{ColangeloReview} P. Colangelo and A. Khodjamirian, hep-ph/0010175.

\bibitem{PDG}  S. Navas et al, Phys. Rev. {\bf D110} (2024) 030001.

\bibitem{Narison-mix} S. Narison and R. Tarrach, Phys. Lett. {\bf 125 B} (1983) 217.

\bibitem{WangHuang3900}  Z. G. Wang and T. Huang,  Phys. Rev. {\bf D89} (2014)  054019.

\bibitem{Wang-tetra-formula}  Z. G. Wang, Eur. Phys. J. {\bf C74} (2014)  2874.

\bibitem{WangZG-mole-formula-1} Z. G. Wang and T. Huang, Eur. Phys. J. {\bf C74} (2014)  2891.

\bibitem{WangZG-mole-formula-2}  Z. G. Wang,  Eur. Phys. J. {\bf C74} (2014)  2963.


\bibitem{Wang-tetra-NPA-2014} Z. G. Wang and T. Huang, Nucl. Phys. {\bf A930} (2014) 63.

\bibitem{Penta-NLO-Groote-PRD-2006} S. Groote, J. G. Korner and A. A. Pivovarov,  Phys. Rev. {\bf D74} (2006) 017503.

\bibitem{Penta-NLO-Groote-PRD-2012} S. Groote, J. G. Korner and A. A. Pivovarov,   Phys. Rev. {\bf D86} (2012) 034023.

\bibitem{WangZG-Landau-PRD} Z. G. Wang, Phys. Rev. {\bf D101} (2020) 074011.

\end{thebibliography}
\end{document}